\documentclass[11pt]{article} 
\pdfoutput=1
\usepackage{amsmath,amssymb,amsfonts,cite,fullpage,graphicx,multirow,nicefrac,slashed}
\usepackage{color}
\definecolor{colorLink}{rgb}{0.9,0,0} 
\definecolor{colorCite}{rgb}{0,0.7,0} 
\definecolor{colorURL} {rgb}{0,0,0.8} 
\usepackage[colorlinks=true,linktocpage=true,linkcolor=colorLink,citecolor=colorCite,urlcolor=colorURL]{hyperref}

\newcommand{\one}          {\mathbf{1}}
\newcommand{\five}         {\mathbf{5}}
\newcommand{\fivebar}      {\overline{\mathbf{5}}}
\newcommand{\ten}          {\mathbf{10}}
\newcommand{\tenbar}       {\overline{\mathbf{10}}}
\newcommand{\fifteen}      {\mathbf{15}}
\newcommand{\fifteenbar}   {\overline{\mathbf{15}}}
\newcommand{\twentyfour}   {\mathbf{24}}
\newcommand{\thirtyfive}   {\mathbf{35}}
\newcommand{\fourtyfour}   {\mathbf{44}}

\newcommand{\fiftyfive}    {\mathbf{55}}
\newcommand{\sixtyfive}    {\mathbf{65}}
\newcommand{\seventyeight} {\mathbf{78}}
\newcommand{\cL}           {\mathcal{L}}
\newcommand{\cO}           {\mathcal{O}}
\newcommand{\Gsm}          {({\rm SU}(3)_c, {{\rm SU}(2)_L)}_{Y}}
\newcommand{\bnum}         {\mathsf{B}}
\newcommand{\lnum}         {\mathsf{L}}

\title{\LARGE\bf Unification and New Particles at the LHC}
\author{\large Nima Arkani-Hamed$^a$, Raffaele Tito D'Agnolo$^a$, Matthew Low$^a$, David Pinner$^b$}
\date{\it \small $^a$School of Natural Sciences, Institute for Advanced Study, Princeton, NJ 08540, USA\\ 
  \vspace{0.8em} $^b$Princeton Center for Theoretical Science, Princeton University, Princeton, NJ 08544, USA}

\begin{document}
\begin{titlepage}
\maketitle
\begin{abstract}
\noindent
Precision gauge coupling unification is one of the primary quantitative
successes of low energy or split supersymmetry.  Preserving this
success puts severe restrictions on possible matter and gauge sectors
that might appear at collider-accessible energies. In this work we
enumerate new gauge sectors which are compatible with unification, consisting of
horizontal gauge groups acting on vector-like matter 
charged under the Standard Model.
Interestingly, almost all of these theories are  in the supersymmetric
conformal window at high energies and confine quickly after the
superpartners are decoupled. For a range of scalar masses compatible
with both moderately tuned and minimally split supersymmetry, the confining
dynamics happen at the multi-TeV scale, leading to a spectrum of multiple
spin-0 and spin-1 resonances accessible to the LHC, with unusual
quantum numbers and striking decay patterns.
\end{abstract}
\end{titlepage}

\tableofcontents

\section{Introduction} \label{sec:introduction}

Direct and indirect searches for new physics at the weak scale have
produced only null results so far. It is looking increasingly more
likely that the widely shared and deeply rooted belief in a perfectly
natural explanation for the origin of the electroweak scale is not
correct, just as we have not seen a natural solution to the
cosmological constant problem.  On the other hand, belief in
naturalness, especially in the context of low-energy supersymmetry (SUSY),
was associated with a spectacular quantitative success -- supersymmetric
gauge coupling unification~\cite{Dimopoulos:1981zb}\footnote{See~\cite{Dienes:1996du} for a review of the original literature on grand unified theories.}
 -- as well as a beautiful and calculable
picture for dark matter candidates~\cite{Jungman:1995df}. If naturalness was far wide of the
mark, why did it lead to these remarkable successes? It is of course
possible that supersymmetric unification is a $\sim 1 \%$ accident,
but it seems more interesting and productive to think about whether
{\it all} of the clues we have seen, both positive and negative, might
be correct, but teaching us a different lesson than we imagined.

One obvious possibility is that SUSY solves the biggest part of the
hierarchy problem, but that a $\sim 1 \%$ level tuning just happens; we
can point to examples of accidental tunings
at this level in other areas of physics. If this is the case, some or all of the superpartners
may not be accessible to the LHC. This is admittedly an uncomfortably
``middling'' possibility -- once tuning is allowed, why stop at the
percent level? Furthermore, moderately heavy $\sim 10$ TeV
superpartners still suffer from flavor and CP problems that require
some degree of model-building gymnastics to avoid.

Minimal split SUSY offers a more coherent picture~\cite{Wells:2003tf,ArkaniHamed:2004fb,Giudice:2004tc,ArkaniHamed:2006mb,Arvanitaki:2012ps,ArkaniHamed:2012gw}. 
From the top down, 
the observation that the most straightforward ways of breaking SUSY
lead to scalars a loop factor heavier than gauginos is already a
strong theoretical motivation for considering a specific level of
split spectrum. The fermion spectrum is anchored by getting the
correct dark matter abundance. The $\sim 10^{-4} - 10^{-6}$
tunings needed for generating the weak scale, which would seem absurd from a
standard ``mono-vacuum'' perspective, seem inconsequential relative
to the vastly larger tunings associated with the cosmological
constant. Furthermore, scalars in the $100-1000$ TeV range are much
more easily compatible with the absence of flavor changing neutral currents (FCNC), CP violation~\cite{ArkaniHamed:2004yi}, and
the cosmological moduli problems~\cite{Banks:1993en,deCarlos:1993wie}. Quite nicely, this
setup predicts a Higgs mass between $\sim 120-135$ GeV~\cite{Bagnaschi:2014rsa}, in agreement
with its observed value.  It is possible, however, that even if split SUSY is
correct, we do not see the fermionic superpartners at the LHC. To take a simple
limit, the WIMP miracle for pure winos and higgsinos needs them to
be near $\sim 3$ TeV and $\sim 1$ TeV respectively. These are too
heavy to be directly produced at the LHC (though they should be
accessible to a possible 100 TeV collider~\cite{Low:2014cba}), and if these masses are at
the bottom of the spectrum, it is quite conceivable that the gluino is
too heavy to be LHC accessible as well.

Thus in both of these pictures it is quite possible that we will be
unlucky and the LHC will not produce the particles motivated to be
near the TeV scale -- either all the superpartners (motivated by
naturalness in conventional SUSY) or the gauginos/higgsinos (motivated by dark
matter in minimal split SUSY). Given this state of affairs, it is
important to ask whether we might get lucky in some other way and
find other new particles which are not directly associated with naturalness or
dark matter, but which might nonetheless show up at the TeV scale.
This is what we will investigate in this paper.

Very naively, there seems to be a vast theoretical zoo of
``unmotivated" new theories and associated new particles
to examine and look for. However, we will find that demanding that any
new particles not spoil the success of supersymmetric gauge coupling
unification puts
severe restrictions on possible matter and gauge sectors that might
appear at collider-accessible energies. We can have new vector-like
matter beyond the minimal (split) supersymmetric particle content. To
automatically preserve unification, the particles should come in multiplets of SU(5),
and there should not be too many of these multiplets at the TeV scale.

The case where the new particles are not charged under strong gauge
dynamics is simple and familiar. The vector-like masses must be near a
TeV, a coincidence that might be explained by linking them to
whatever explains vector-like higgsino masses. The associated
phenomenology is also familiar, consisting of vector-like
$(\five + \fivebar)$s and $(\ten + \tenbar)$s (but no larger multiplets) that may appear as
collider-stable charged and colored particles or decay by mixing with
standard-model fermions (see~\cite{Aguilar-Saavedra:2013qpa} and references therein).

The case where the vector-like matter is charged under new strong gauge
dynamics is more interesting~\cite{Kilic:2009mi}. We enumerate all new gauge groups that
can act on vector-like matter, finding a surprisingly small list that
can be explored systematically. Interestingly, almost all of these
theories are in the supersymmetric conformal window at high energies~\cite{Seiberg:1994pq,Intriligator:1995ne}
and confine quickly after superpartners are decoupled. For a wide
range of scalar masses compatible with both mildly-tuned and
minimally split SUSY, confinement happens at the multi-TeV scale,
giving rise to a panoply of still-lighter ``pions'' and ``$\rho$ mesons''
charged under the Standard Model (SM) gauge interactions.  Thus, if these additional
gauge groups are present, even if the particles motivated by naturalness or dark-matter 
are not in reach, a rich (but systematically
classifiable) spectrum of new particles with striking experimental
signatures could be accessible to the LHC.

\section{Unification and the Conformal Window} \label{sec:theories}

\begin{table}\begin{center}
\begin{tabular}{ccc}
confining group $G_H$ & flavor symmetry                              & $~~N_F~~$ \\ \hline
SU(2)$_H$             & SU($2N_F$) $\to$ Sp($2N_F$)                  & $\leq$ 6  \\
SU(3)$_H$             & SU($N_F$) $\times$ SU($N_F$) $\to$ SU($N_F$) & $\leq$ 9  \\
SU(4)$_H$             & SU($N_F$) $\times$ SU($N_F$) $\to$ SU($N_F$) & $\leq$ 12 \\
Sp(4)$_H$             & SU($2N_F$) $\to$ Sp($2N_F$)                  & $\leq$ 9
\end{tabular}
\caption{Asymptotically free horizontal gauge groups compatible with perturbative unification.  SM gauge interactions are contained in a SU(5) subgroup of the flavor symmetry.}
\label{tab:gaugegroups}
\end{center}\end{table}

There is a short list of models with strong dynamics which are consistent with precision gauge coupling unification.  As we will show, the confinement scale is tied to the superpartner mass scale in almost all cases.

In order to preserve unification, we require all additional matter that is charged under the SM to transform in complete GUT multiplets.  For the unified coupling to remain perturbative up to the GUT scale, $\alpha_{\rm GUT} (M_{\rm GUT}) \lesssim 0.3$, there can be no more than four $\five$ + $\fivebar$ pairs or a single $\ten$ + $\tenbar$ pair that can be accompanied by a single $\five$ + $\fivebar$ pair~\cite{Masip:1998jc,Jones:2008ib,Martin:2010kk}.  The flavor symmetry which exchanges these vector-like pairs, which we call $G_H$ (sometimes referred to as hypercolor~\cite{Kilic:2009mi}), commutes with SU(5) and can therefore be gauged consistent with SM symmetries.  In general, SU(5) can be a subgroup of a larger (partially gauged) global flavor symmetry, so that the new matter transforms under $G = G_H \times G_F$, with SU(5) $\subset G_F$.

In this setting, constraining the SM-charged matter content of the theory automatically reduces the possible choices for $G_H$.  We are limited to groups that have representations with dimension smaller than or equal to four. This includes the SU($N$) groups with $N \leq 4$, the SO($N$) groups with $N \leq 6$, and the Sp($2N$) groups with $N \leq 2$.\footnote{There aren't any viable exceptional groups, and the only viable product groups are isomorphic to SO(4).}  Isomorphisms between Lie algebras leave us with SU($N$)$_H$ with $2 \leq N \leq 4$, SO(4)$_H$, and Sp(4)$_H$.  

If we require $G_H$ to be asymptotically free, we can embed the new matter in the fundamental representation of SU($N$)$_H$ with $2 \leq N \leq 4$ or Sp(4)$_H$, while SO(4)$_H$ is ruled out. The number of flavors (with respect to $G_H$) must be at least five if we want to unify into SU(5), but in principle the flavor symmetry can be enlarged without spoiling unification, resulting in additional SM singlets. The maximum number of flavors allowed for each group is listed in Table~\ref{tab:gaugegroups}.

\begin{figure}\begin{center}
\includegraphics[width=0.6\textwidth]{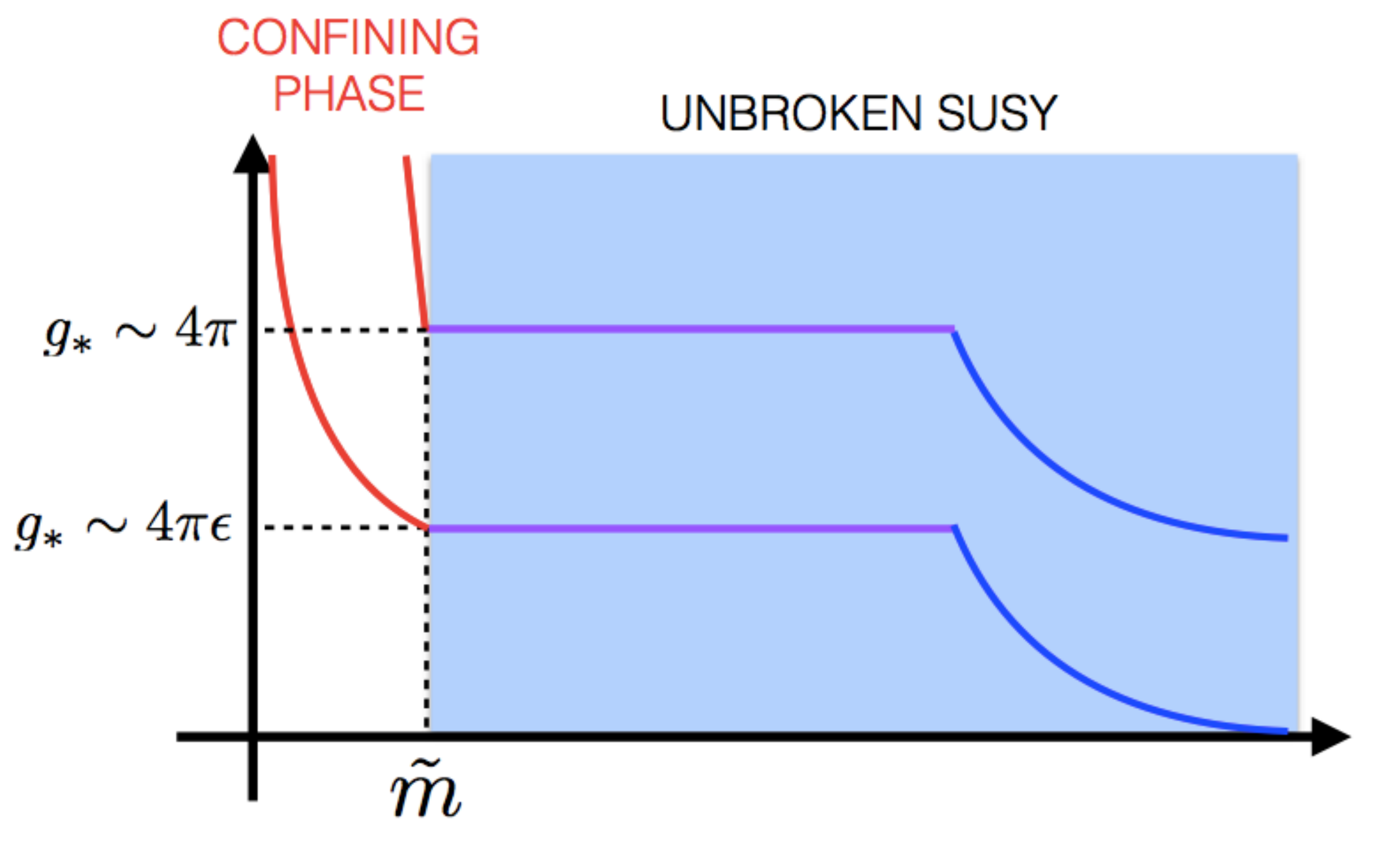}
\caption{Schematic representation of our setup. In the UV the gauge theories in Table~\ref{tab:gaugegroups} are in the supersymmetric conformal window and flow towards their IR fixed point (blue). The fixed point may be reached before the superpartners are decoupled, in which case the theory becomes conformal (purple).  Once the superpartners are decoupled at $\tilde m$, the new gauge interactions enter a confining phase (red).}
\label{fig:cartoon}
\end{center}\end{figure}

The restrictions on $N$ and $N_F$ discussed above have a very interesting consequence: most of the theories in Table~\ref{tab:gaugegroups} are in the supersymmetric conformal window in the UV~\cite{Seiberg:1994pq,Intriligator:1995ne}.  This is shown schematically in Fig.~\ref{fig:cartoon}.  More precisely, SU(2)$_H$, SU(3)$_H$, and Sp(4)$_H$ are in the supersymmetric conformal window in the UV.  SU(4)$_H$ may be in the conformal window as well, but only if $N_F \geq 7$.  Once the superpartners have been decoupled, all of these theories enter a confining phase according to lattice computations~\cite{Karavirta:2011zg,Fodor:2015baa} or analytic estimates~\cite{Ryttov:2007sr,Sannino:2009aw}.  Lattice simulations have shown that for SU(2)$_H$ $N_F=6$ is expected to be close to the boundary of the (non-supersymmetric) conformal window, while $N_F=4$ is confining~\cite{Karavirta:2011zg}.  For SU(3)$_H$ $N_F \leq 8$ is confining~\cite{Fodor:2015baa}.  The other cases have not been computed on the lattice, but analytic methods suggest that confinement occurs for all possible $N_F$ values in Table~\ref{tab:gaugegroups}~\cite{Ryttov:2007sr,Sannino:2009aw}.

How much below the scale $\tilde{m}$ (where the superpartners are decoupled) confinement actually occurs depends on the value of the gauge coupling, $g$, evaluated at $\tilde{m}$.  For example, in SU($N$)$_H$ supersymmetric gauge theories near the top of the conformal window ($N_F = 3 N-\epsilon N$ for $\epsilon \ll 1$), the fixed point is $g_*^2 = \frac{8\pi^2}{3}\frac{N}{N^2-1}\epsilon + \mathcal{O}(\epsilon^2)$, with $g_*$ growing as we move towards the bottom of the window ($N_F \to 3 N/2$). In the theories we discuss, given the values of $N$ and $N_F$ in Table~\ref{tab:gaugegroups}, $g_*$ is always rather large, meaning that confinement generically occurs immediately below $\tilde{m}$, or else a decade or two below. In practice it is hard to say anything more precise, due to the size of the coupling.\footnote{Conceptually it is possible that the theory is still flowing towards the fixed point as we hit $\tilde m$ and that $g(\tilde m)<g_*$.  In this case several decades of running could be needed before confinement occurs.}

We see that, in this class of theories, the confinement scale is tied to the superpartners mass scale. Even abandoning naturalness completely, the observed value of the Higgs mass requires superpartners below $\sim 100-1000$~TeV, unless $\tan\beta$ is extremely close to one, either in split or high scale supersymmetry~\cite{ArkaniHamed:2012gw,Bagnaschi:2014rsa}.  Gauge coupling unification favors a similar upper bound~\cite{Bagnaschi:2014rsa}. As discussed previously, there is also a host of other considerations that point to superpartners in this range, from the absence of FCNC's and CP signals to dark matter in split scenarios.  Thus it is reasonable to expect a confinement scale between $1000$~TeV and a TeV.

Now consider the minimal picture in which this scenario is realized. In the low-energy theory, the flavor symmetry of the confining group is explicitly broken only by SM gauge interactions. Then after confinement and spontaneous chiral symmetry breaking, we are left with pseudo-Goldstone bosons with masses a SM loop factor below the confinement scale.  Even in the phenomenologically worst case, when the theory confines around $1000$~TeV, this can be interesting both for the LHC and a future 100 TeV proton-proton machine.

In the following section we discuss the phenomenology of this class of theories.  For the reasons explained above, we will focus mainly on models with small explicit breaking of the chiral symmetry.  In such cases the spectrum of the theory contains pseudo-Goldstone bosons parametrically lighter than the confinement scale.

\section{Particles and Interactions} \label{sec:pheno}

\subsection{Matter Content}

If explicit chiral symmetry breaking is small, the lightest states in the theory are its pseudo-Goldstone bosons, which from now on we call pions.  For the gauge groups SU(3)$_H$ and SU(4)$_H$, with the minimal number of flavors, the symmetry breaking pattern is SU(5) $\times$ SU(5) $ \to $ SU(5), which results in pions in the $\twentyfour$ representation of the unbroken, diagonal SU(5).  When the gauge group is SU(2)$_H$ or Sp(4)$_H$ we expect the symmetry breaking pattern to be instead SU(10) $\to$ Sp(10) which results in pions in the $\fourtyfour$ of Sp(10)~\cite{Peskin:1980gc}.  In terms of SU(5), these pions decompose as
\begin{equation} \label{eq:decomp44}
\fourtyfour \to \twentyfour + \ten + \tenbar.
\end{equation}
Thus we see that the $\twentyfour$ is common to all cases, but the additional pions in the $\ten$ and $\tenbar$ are only present for the gauge groups SU(2)$_H$ and Sp(4)$_H$.

Since we are always considering theories with weakly coupled UV completions, it is useful to explicitly introduce the $\five$ and $\fivebar$'s that condense to form the pions
\begin{equation}
F^i   = ( D ,\; L^c )^i, \quad\quad\quad
F^c_i = ( D^c ,\; L )_i\, .
\end{equation}
Here $i$ runs from 1 to $N$ which specifies the number of fiveplets added.  The $F_i$ transform as a fundamental of $G_H$, with the $F^c_i$ in the antifundamental. As usual, $D$ is a color triplet with hypercharge $Y=-\nicefrac{1}{3}$, and $L$ is an SU(2)$_L$ doublet with hypercharge $Y=\nicefrac{-1}{2}$.

The pions in the $\twentyfour$ are summarized in Table~\ref{tab:states-pions24} along with their constituents and quantum numbers. There is a color octet, $\pi_8$, an electroweak triplet, $\pi_3$, a SM singlet, $\pi_1$, and an exotic, colored pion $Q_X$ that is both an electroweak doublet and a color triplet.  In addition, there is a second SM singlet, $\pi'$, which we include in the pion list even though it receives mass contributions independent of chiral symmetry breaking, because it mixes with $\pi_1$ after explicit breaking of SU(5).

\begin{table}\begin{center}
\begin{tabular}{ccc}
meson                                                & constituents        & $\Gsm$                                            \\ \hline
$\pi_8$                                              & $D^c D$             & $(\mathbf{8},\mathbf{1})_0$                       \\
$\pi_3$                                              & $L L^c$             & $(\mathbf{1},\mathbf{3})_0$                       \\
$\pi_1$                                              & $2 D^c D - 3 L L^c$ & $(\mathbf{1},\mathbf{1})_0$                       \\
$Q_X = (X_{\nicefrac{-1}{3}}, X_{\nicefrac{-4}{3}})$ & $L D$               & $(\mathbf{3},\mathbf{2})_{\nicefrac{-5}{6}}$      \\
$Q_X^* = (X_{\nicefrac{4}{3}}, X_{\nicefrac{1}{3}})$ & $D^c L^c$           & $(\mathbf{\bar{3}},\mathbf{2})_{\nicefrac{5}{6}}$ \\
$\pi'$                                               & $D^c D + L L^c$     & $\mathbf{1}$ 
\end{tabular}
\caption{Vector-like quark constituents and SM gauge groups representations of the pions in the $\twentyfour$ of SU(5).}
\label{tab:states-pions24}
\end{center}\end{table}

There are two sources of explicit breaking of SU(5) $\times$ SU(5) (or SU(10)) that give mass to the pions: loops of SM gauge bosons and vector-like constituent masses~\cite{Kilic:2009mi}.\footnote{States in electroweak multiplets are split by $\sim \alpha M_Z$, but this is not relevant for the phenomenology we discuss.}   The gauge loop contribution is fixed by the pion representations under the SM groups and gives a lower bound on their mass
\begin{equation}
\delta m_\pi^2 = \frac{3c^2 \Lambda^2}{16 \pi^2} \sum_{G} g_G^2 C_2^G(\pi).
\end{equation}
Here $C_2^G(\pi)$ is the quadratic Casimir of the pion representation under $G=\{ {\rm SU(3)}_c, {\rm SU(2)}_L, {\rm U(1)}_Y \}$, and $\Lambda$ is the confinement scale of the new strong interactions.  The parameter $c$ represents an unknown coefficient, as the quadratically-divergent loops are cut off by particles from the strongly-interacting sector.  In the following we set $c=1$, which corresponds to assuming that the loop is cut off by vector meson resonances with masses comparable to the confinement scale.

As in the Standard Model, explicit breaking of the chiral symmetry gives the pions mass at tree level.  There are two parameters, $M_D$ and $M_L$, which are invariant under the SM gauge groups.  If these are equal at the GUT scale, then at low energies $M_D \approx 2 M_L$.  While it is possible to stray from this relation, deviations would indicate GUT-breaking boundary conditions (or threshold corrections).

\begin{figure}\begin{center}
\includegraphics[width=0.45\textwidth]{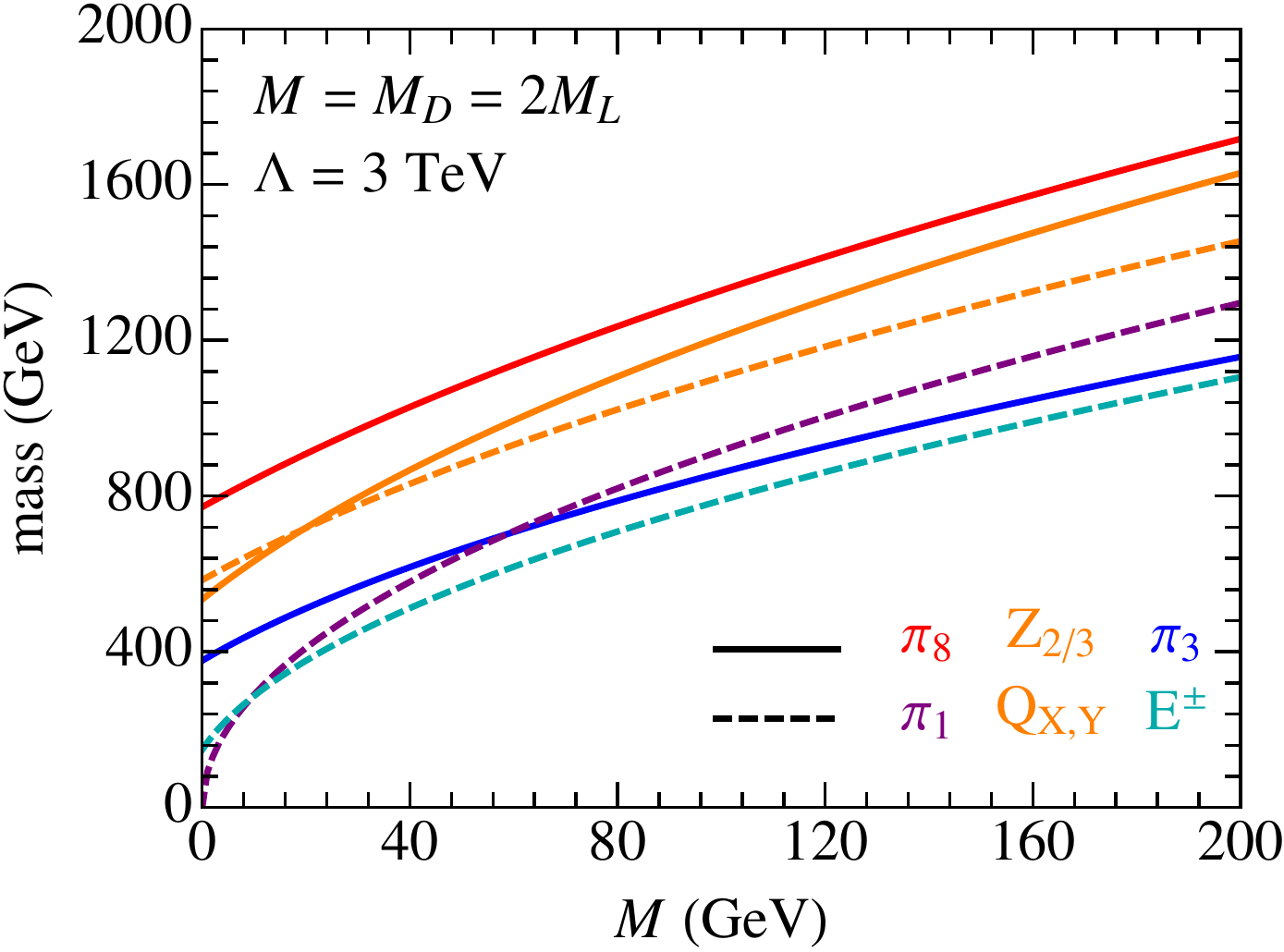}
\caption{Masses of the pions in the $\fourtyfour$ of Sp(10) as a function of the vector-like mass $M=M_D=2 M_L$.}
\label{fig:pions44-masses}
\end{center}\end{figure}

The pion masses, including both contributions, are shown in Fig.~\ref{fig:pions44-masses} as a function of the vector-like quark masses for $\Lambda=3$~TeV. At leading order in $M_D$ and $M_L$ they are
\begin{subequations}\label{eq:24Masses}\begin{align}
m_{\pi_1}^2 &= 4\Lambda \frac{2M_D + 3M_L}{5}, \\
m_{\pi_3}^2 &= 4\Lambda M_L +\frac{3 g^2 \Lambda^2}{8\pi^2}, \\
m_{Q_X}^2   &= 4\Lambda \frac{M_D + M_L}{2} + \frac{g_3^2\Lambda^2}{4\pi^2}+\frac{9g^2\Lambda^2}{64\pi^2}+\frac{25g'^2\Lambda^2}{192\pi^2}, \\
m_{\pi_8}^2 &= 4\Lambda M_D +\frac{9g_3^2\Lambda^2}{16\pi^2}.
\end{align}\end{subequations}
Here we have used the naive dimensional analysis (NDA) estimate for the fermion condensate, $\langle F^c_i F^j \rangle = f^2 \Lambda {\delta_i}^j$, and chosen a normalization corresponding to $f=92$~MeV in QCD~\cite{Georgi:1992dw}. 

The additional pions in the $\fourtyfour$ are listed in Table~\ref{tab:states-pions44}.  These include another color triplet and electroweak doublet $Q_Y$, which is almost mass-degenerate with $Q_X$, a color triplet $Z_{\nicefrac{2}{3}}$, and a charged pion $E^\pm$ that is close in mass to $\pi_3$.  Their masses are also shown in Fig.~\ref{fig:pions44-masses}, and at leading order in $M_D$ and $M_L$ they read

\begin{table}\begin{center}
\begin{tabular}{ccc}
meson                                                 & constituents & $\Gsm$                                             \\ \hline
$Q_Y = (Y_{\nicefrac{2}{3}}, Y_{\nicefrac{-1}{3}})$   & $L^c D$      & $(\mathbf{3},\mathbf{2})_{\nicefrac{1}{6}}$        \\
$Q_Y^* = (Y_{\nicefrac{1}{3}}, Y_{\nicefrac{-2}{3}})$ & $D^c L$      & $(\mathbf{\bar{3}},\mathbf{2})_{\nicefrac{-1}{6}}$ \\
$Z_{\nicefrac{2}{3}}$                                 & $D^c D^c$    & $(\mathbf{3},\mathbf{1})_{\nicefrac{2}{3}}$        \\
$Z_{\nicefrac{2}{3}}^*$                               & $D D$        & $(\mathbf{\bar{3}},\mathbf{1})_{\nicefrac{-2}{3}}$ \\
$E^+$                                                 & $L^c L^c$    & $(\mathbf{1},\mathbf{1})_1$                        \\
$E^-$                                                 & $L L$        & $(\mathbf{1},\mathbf{1})_{-1}$               
\end{tabular}
\caption{Pions in the $\ten$ and $\tenbar$ of SU(5) that appear in the theories with a Sp(2$N_F$) flavor symmetry.}
\label{tab:states-pions44}
\end{center}\end{table}

%
\begin{subequations}\label{eq:44Masses}\begin{align}
m_{Q_Y}^2                 &= 4\Lambda \frac{M_D + M_L}{2} + \frac{g_3^2 \Lambda^2}{4\pi^2}+\frac{9g^2\Lambda^2}{64\pi^2}+\frac{g'^2\Lambda^2}{192\pi^2}, \\
m_{Z_{\nicefrac{2}{3}}}^2 &= 4\Lambda M_D + \frac{g_3^2\Lambda^2}{4\pi^2} + \frac{g'^2\Lambda^2}{12\pi^2}, \\
m_{E^\pm}^2               &= 4\Lambda M_L + \frac{3g'^2\Lambda^2}{16\pi^2}\, .
\end{align}\end{subequations}

This completes the set of possibilities for $N_F = 5$.  However, it is only the matter charged under the SM that is limited by perturbative unification. As discussed in Sec.~\ref{sec:theories} we can add new SM singlets and enlarge the flavor symmetry. So for $N_F > 5$ we expect new $\five$, $\fivebar$, and singlet pions in addition to those already present for $N_F = 5$. To see this in more detail, let us consider the case $N_F = 6$.

The constituent fields are extended to
\begin{equation}
F^i   = ( D ,\; L^c, \; S )^i, \quad\quad\quad
F^c_i = ( D^c ,\; L, \; S^c )_i,
\end{equation}
where $S$ is a SM singlet.  Furthermore, there is an additional relevant parameter, $M_S$, which explicitly breaks SU(6) $\times$ SU(6) and contributes to the pion masses.  The pions, in this case, are in the $\thirtyfive$ of SU(6).  Under the SU(5) subgroup containing the SM gauge groups, this representation decomposes as
\begin{equation} \label{eq:decomp35}
\thirtyfive \to \twentyfour + \five + \fivebar + \one.
\end{equation}
Thus, in addition to the usual $\twentyfour$, there are also a composite $\five$ + $\fivebar$ and a singlet.  These states are listed in Table~\ref{tab:states-pions35}.  The masses of $\phi_D$ and $\phi_L^*$ (the new states in the $\five$) are
\begin{subequations}\label{eq:SU6Masses}\begin{align}
m_{\phi_D}^2 &= 4\Lambda \frac{M_D + M_S}{2} + \frac{g_3^2\Lambda^2}{4\pi^2} + \frac{g'^2\Lambda^2}{48\pi^2}, \\
m_{\phi_L}^2 &= 4\Lambda \frac{M_L + M_S}{2} + \frac{9g^2\Lambda^2}{64\pi^2} + \frac{3g'^2\Lambda^2}{64\pi^2}.
\end{align}\end{subequations}

As in the case of SU(5) $\times$ SU(5), the singlet states can mix with $\pi'$.  This mixing, which is proportional to SU(5) and SU(6) violation for $\pi_1$ and $\pi_S$, respectively, vanishes in the chiral limit.  However, $\pi_S$ and $\pi_1$ still mix with one another in the presence of SU(5) violation.  In this case the singlet mass eigenstates are
\begin{subequations}\begin{align}
a_1 &= \pi_1 \cos\theta_S + \pi_S \sin\theta_S, \\
a_2 &= \pi_S \cos\theta_S - \pi_1 \sin\theta_S, \\
\tan 2\theta_S &= \frac{12 (M_D - M_L)}{9 M_D + 16 M_L - 25 M_S},
\end{align}\end{subequations}
with masses
\begin{equation}
m_{a_{1,2}}^2 = \frac{\Lambda}{3} \left(4 M_L + 3 M_D + 5 M_S \pm \sqrt{16 (M_L - M_S)^2 + 9 (M_D - M_S)^2} \right).
\end{equation}

\begin{table}\begin{center}
\begin{tabular}{ccc}
meson                                     & constituents   & $\Gsm$                                            \\ \hline
$\phi_D$                                  & $D S^c$        & $(\mathbf{3},\mathbf{1})_{\nicefrac{-1}{3}}$      \\
$\phi_D^*$                                & $S D^c$        & $(\mathbf{\bar{3}},\mathbf{1})_{\nicefrac{1}{3}}$ \\
$\phi_L = ( \phi^0_L ,\; \phi^-_L )$      & $S L$          & $(\mathbf{1},\mathbf{2})_{\nicefrac{-1}{2}}$      \\
$\phi_L^* = ( \phi^+_L ,\; \phi^{0*}_L )$ & $L^c S^c$      & $(\mathbf{1},\mathbf{2})_{\nicefrac{1}{2}}$       \\
$\pi_S$                                   & $S S^c$        & $(\mathbf{1},\mathbf{1})_0$                    
\end{tabular}
\caption{Pions in the $\five$, $\fivebar$, and singlet representations of SU(5) contained in the $\thirtyfive$ of SU(6).}
\label{tab:states-pions35}
\end{center}\end{table}

Finally, when the confining group is SU(2)$_H$ or Sp(4)$_H$, the symmetry breaking pattern is SU(12) $\to$ Sp(12).  The pions are in a $\sixtyfive$ of Sp(12), which decomposes under SU(5) as
\begin{equation} \label{eq:decomp65}
\sixtyfive \to \twentyfour + \ten +\tenbar + 2 \times (\five + \fivebar) + \one.
\end{equation}
These are all representations that we have already discussed, including the $\five$ + $\fivebar$ pairs and the $\one$ that are present in SU(6) $\times$ SU(6).  The new mesons are an extra $\five$ + $\fivebar$ (with constituents $DS$ and $L^c S$), and their phenomenology is essentially the same as for the mesons in SU(6) $\times$ SU(6).  For even larger global symmetries, very little new phenomenology appears with respect to $N_F = 5$ or 6.  Let the global symmetry be $N_F = 5 + \Delta$.  For SU($N_F$) $\times$ SU($N_F$) there are $\Delta$ additional $\five$ + $\fivebar$ pairs and $\Delta^2$ additional singlets.  In SU($2N_F$) there are $2\Delta$ additional $\five$ + $\fivebar$ pairs and $\Delta(2\Delta-1)$ new singlets.

In addition to pions, spin-1 states may be phenomenologically relevant, even with masses of $\mathcal{O}(\Lambda)$.  There are both vectors, which couple to the currents of the unbroken global symmetry, and axial vectors, which couple to the currents of the broken global symmetry.  For the gauge groups SU(3)$_H$ and SU(4)$_H$, the vectors fall into the $\twentyfour$ + $\one$ of SU(5) and the axial vectors into the $\twentyfour$.  

When the gauge group is SU(2)$_H$ or Sp(4)$_H$, the vector mesons now come in the $\fiftyfive$ of Sp(10) and the axial vectors in the $\fourtyfour$ of Sp(10).  The axial vectors in the $\fourtyfour$ decompose under SU(5) according to Eq.~\eqref{eq:decomp44}, while the vectors in the $\fiftyfive$ decompose as
\begin{equation}
\fiftyfive \to \twentyfour + \fifteen + \fifteenbar + \one.
\end{equation}
Again, the $\twentyfour$ is common to all cases, but we have additional states in the $\fifteen$ and $\fifteenbar$.  These vector multiplets contain particles with the SM quantum numbers
\begin{equation}
\fifteen \to 
(\mathbf{3}, \mathbf{2})_{\nicefrac{1}{6}}
+ (\mathbf{1}, \mathbf{3})_{1}
+ (\mathbf{6}, \mathbf{1})_{\nicefrac{-2}{3}} .
\end{equation}

As for the pions, going to $N_F>5$ introduces new $\five$s, $\fivebar$s and singlets.  For the SU(6) $\times$ SU(6) global symmetry the vectors are in the $\thirtyfive + \one$ of SU(6) and the axial vectors are in the $\thirtyfive$ of SU(6).  These decompose according to Eq.~\eqref{eq:decomp35}.  For the SU(12) global symmetry the vectors are in the $\seventyeight$ of Sp(12) while the axial vectors are in the $\sixtyfive$ of Sp(12).  Under SU(5) the $\sixtyfive$ decomposition is given in Eq.~\eqref{eq:decomp65} and the decomposition of the $\seventyeight$ is
\begin{equation}
\seventyeight \to \twentyfour + \fifteen + \fifteenbar + 2 \times (\five + \fivebar) + 4 \times (\one).
\end{equation}

As mentioned above, we are just adding more $\five$s, $\fivebar$s and singlets compared to the $N_F=5$ case and the same happens for $N_F >6$.

\subsection{Pion Phenomenology} \label{sec:PiPheno}

\begin{figure}\begin{center}
\includegraphics[width=0.6\textwidth]{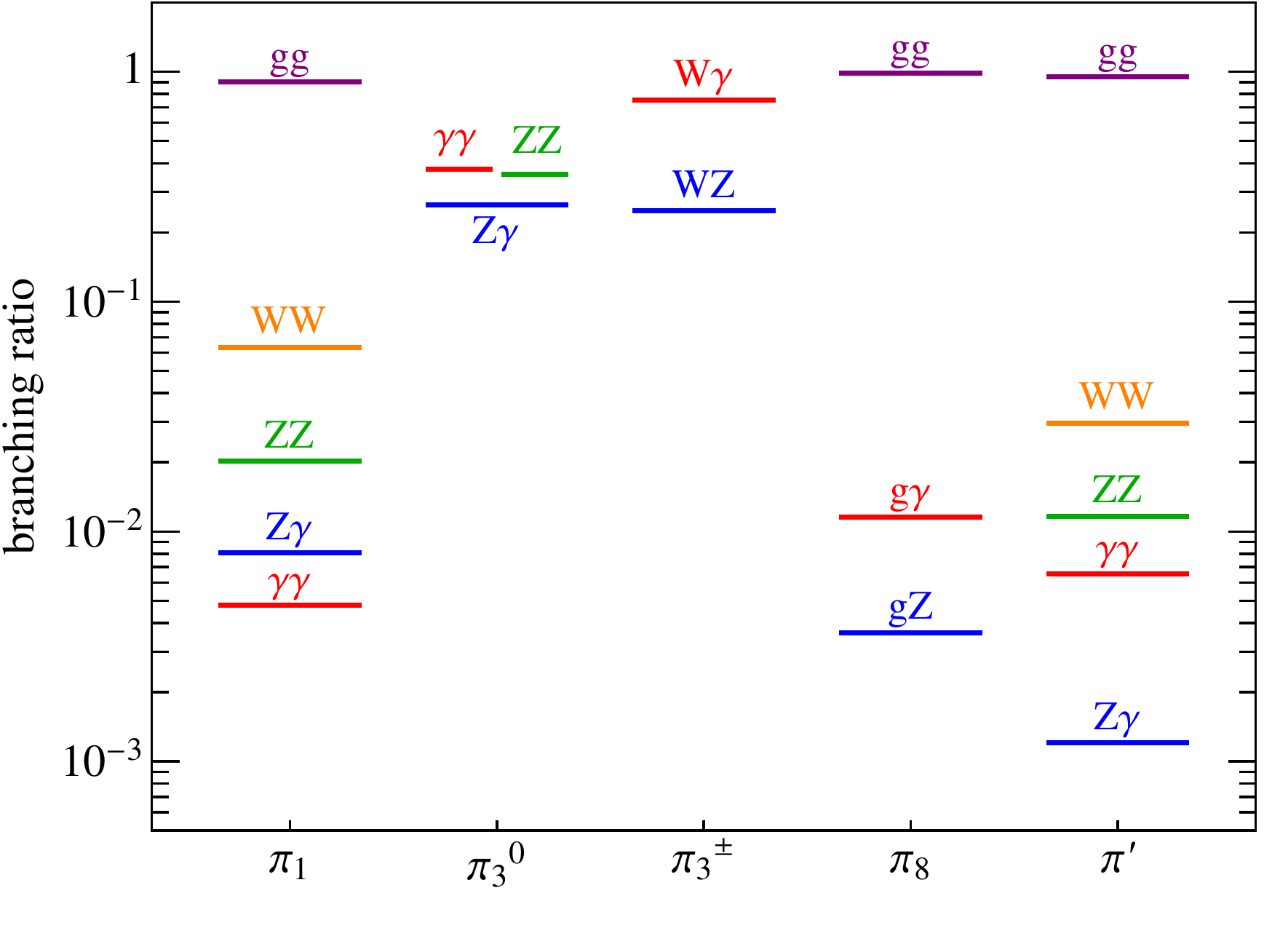}
\caption{Branching ratios of the real pions contained in the $\twentyfour$ of SU(5). In this figure the SM gauge couplings are evaluated at 1 TeV.}
\label{fig:pions24-brs}
\end{center}\end{figure}

It is useful to distinguish between real pions, {\it i.e.}, those that are in real representations of the SM gauge groups, and complex pions, which are not.  All of the real pions come from the $\twentyfour$ of SU(5) and are present in all the theories that we have discussed.  For SU(3)$_H$ and SU(4)$_H$ with $N_F=5$, the only complex pion is $Q_X$, while for SU(2)$_H$ and Sp(4)$_H$ there are also $Q_Y$, $Z_{\nicefrac{2}{3}}$, and $E^\pm$.  Finally, when $N_F > 5$ there are the additional complex pions $\phi_D$ and $\phi_L$.

Real pions interact with pairs of SM vectors through the global SU(5) anomaly 
\begin{equation}
\cL \supset -\frac{N g_A g_B}{16\pi^2 f} \epsilon_{\mu\nu\alpha\beta} \text{tr}(\pi F^{\mu\nu}_A F^{\alpha\beta}_B),
\end{equation}
where $A,B \in \{$SU(3)$_c$, SU(2)$_L$, U(1)$_Y \}$, $f$ is the chiral symmetry breaking scale, and $\pi$, $F_A$, and $F_B$ are all embedded in SU(5).  Through this coupling, the real pions always decay promptly.  The branching ratios are fixed by their gauge quantum numbers and are shown in Fig.~\ref{fig:pions24-brs}.\footnote{Note that the real pions branching ratios weakly depend on their masses through the running of the SM gauge couplings. For example the branching ratios of $\pi_8$, $\pi_1$, and $\pi'$ to two electroweak gauge bosons vary by $\approx 10\%$ when changing the masses between 1 and 2 TeV.}

\begin{figure}\begin{center}
\includegraphics[width=0.45\textwidth]{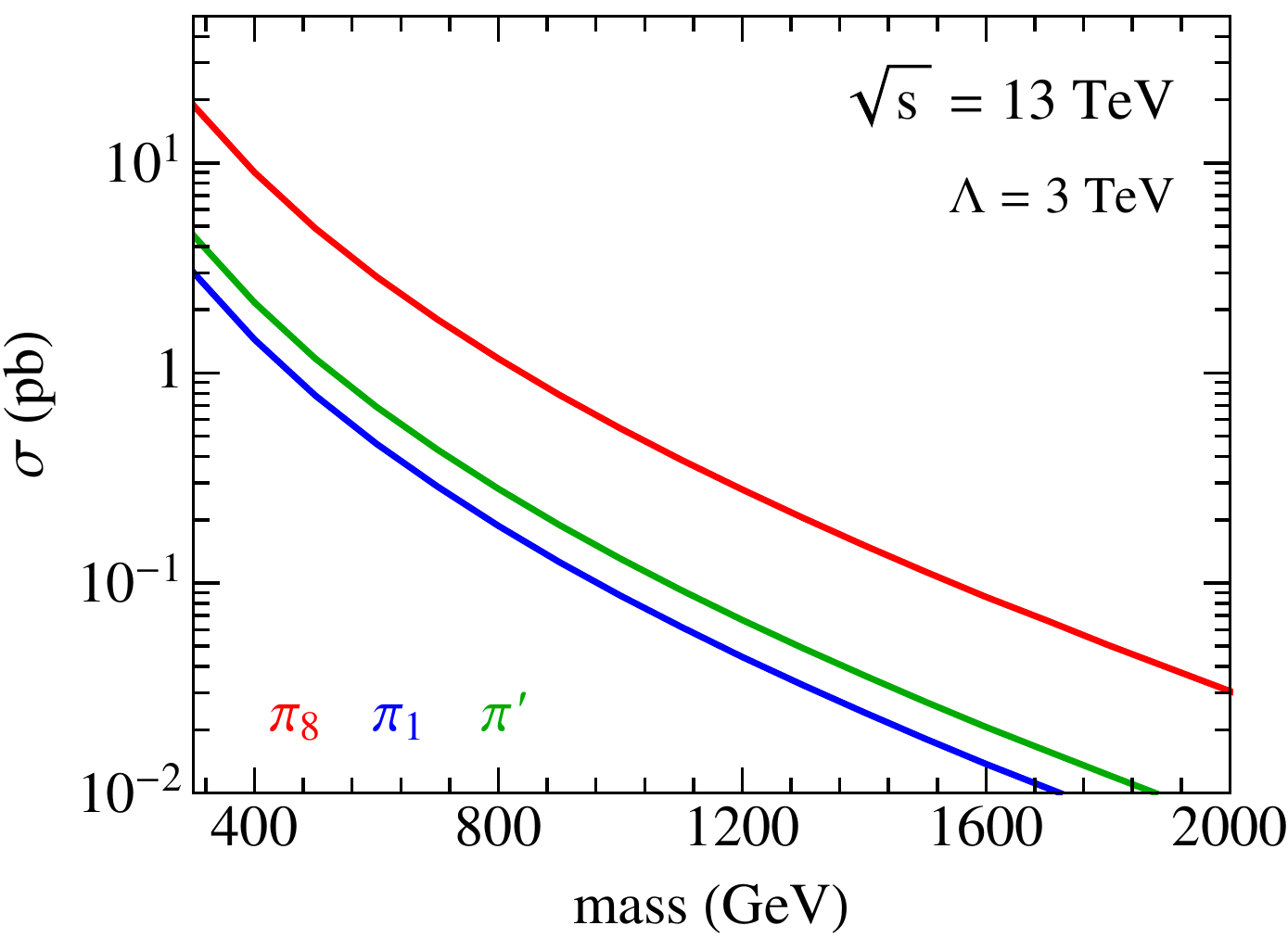} \quad\quad
\includegraphics[width=0.43\textwidth]{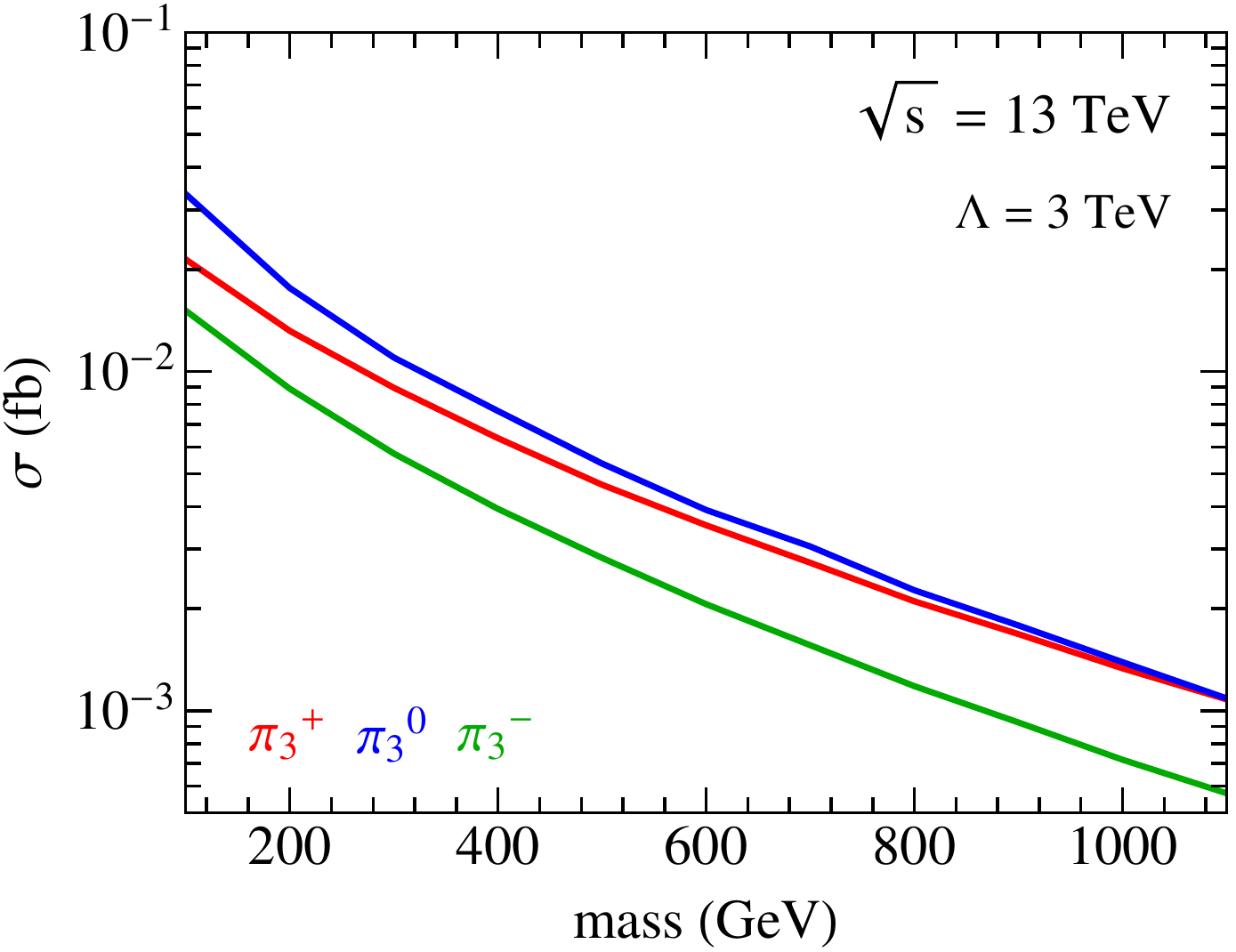}
\caption{The single production cross section of real pions in gluon fusion (left) and the single production cross section of pions in vector boson fusion (right). Both figures use $N=2$ and $\Lambda=$ 3 TeV and assume a single strong coupling in the confining sector, $g_\rho=4\pi/\sqrt{2}$.}
\label{fig:pions24-xsec-single}
\end{center}\end{figure}

The coupling to the anomaly also leads to single production of $\pi_8$, $\pi_1$, and $\pi'$ in gluon fusion.  Although the rate, which is proportional to $N^2/f^2$ (or $N/\Lambda^2$), may seem highly model-dependent, recall there is a finite list of choices for $N$ (see Table~\ref{tab:gaugegroups}), and $\Lambda$ is well-motivated to be near the TeV scale.  The left panel of Fig.~\ref{fig:pions24-xsec-single} shows the single production rate in gluon fusion using $N=2$ and $\Lambda=$ 3 TeV.  The right panel shows the single production of electroweak pions in vector boson fusion.

From single production of $\pi_8$, one expects to see resonances in the dijet, $j\gamma$, and $jZ$ spectra.  Similarly, there should be resonances in the dijet, $WW$, $ZZ$, $Z\gamma$, and $\gamma\gamma$ channels at the $\pi_1$ mass.  $\pi_3$ decays lead to $W\gamma$ and $WZ$ resonances in vector boson fusion, albeit at much lower rates.  Other signatures will be discussed in more detail in Sec.~\ref{sec:signals}.

\begin{figure}\begin{center}
\includegraphics[width=0.45\textwidth]{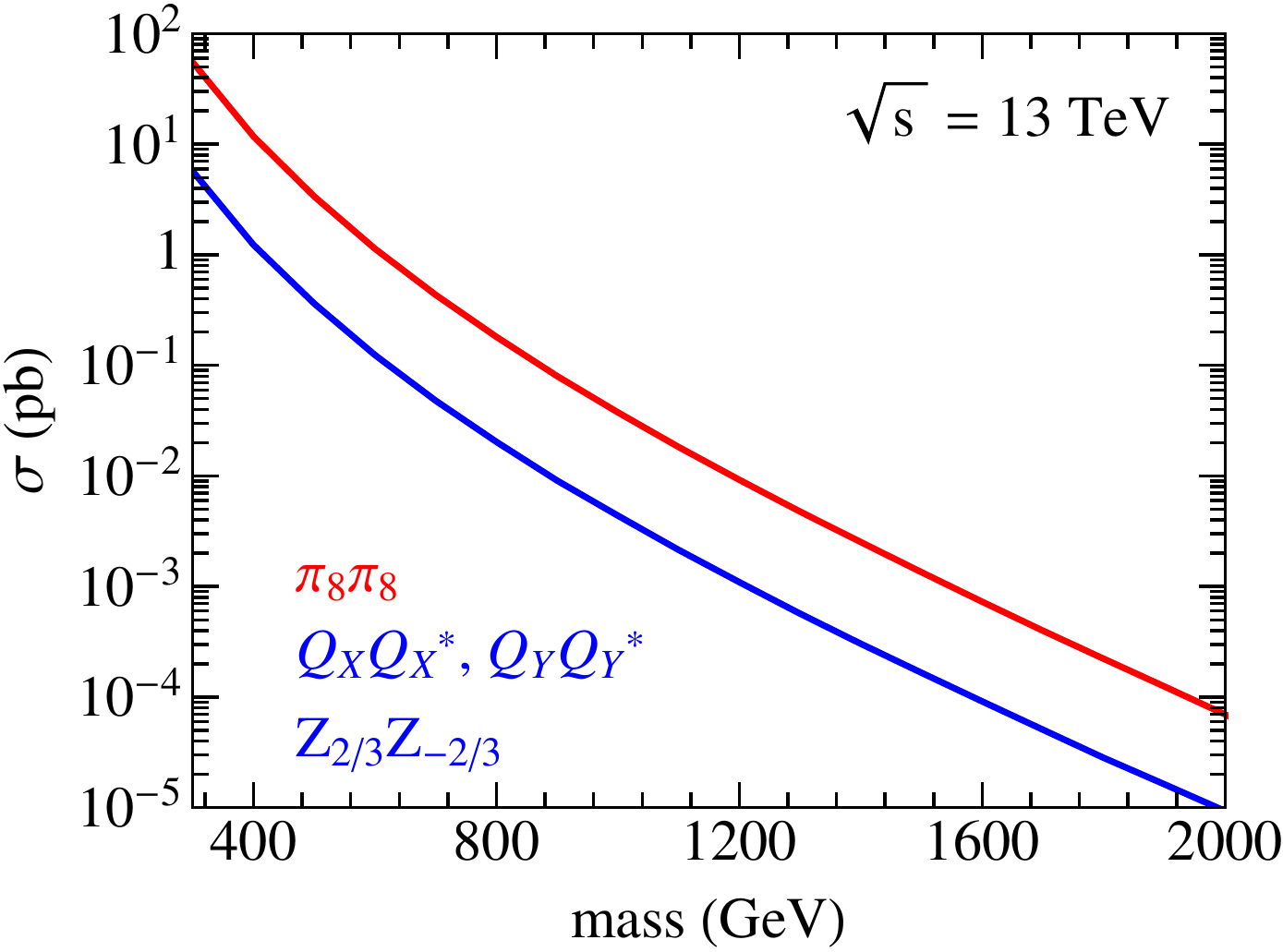} \quad\quad
\includegraphics[width=0.45\textwidth]{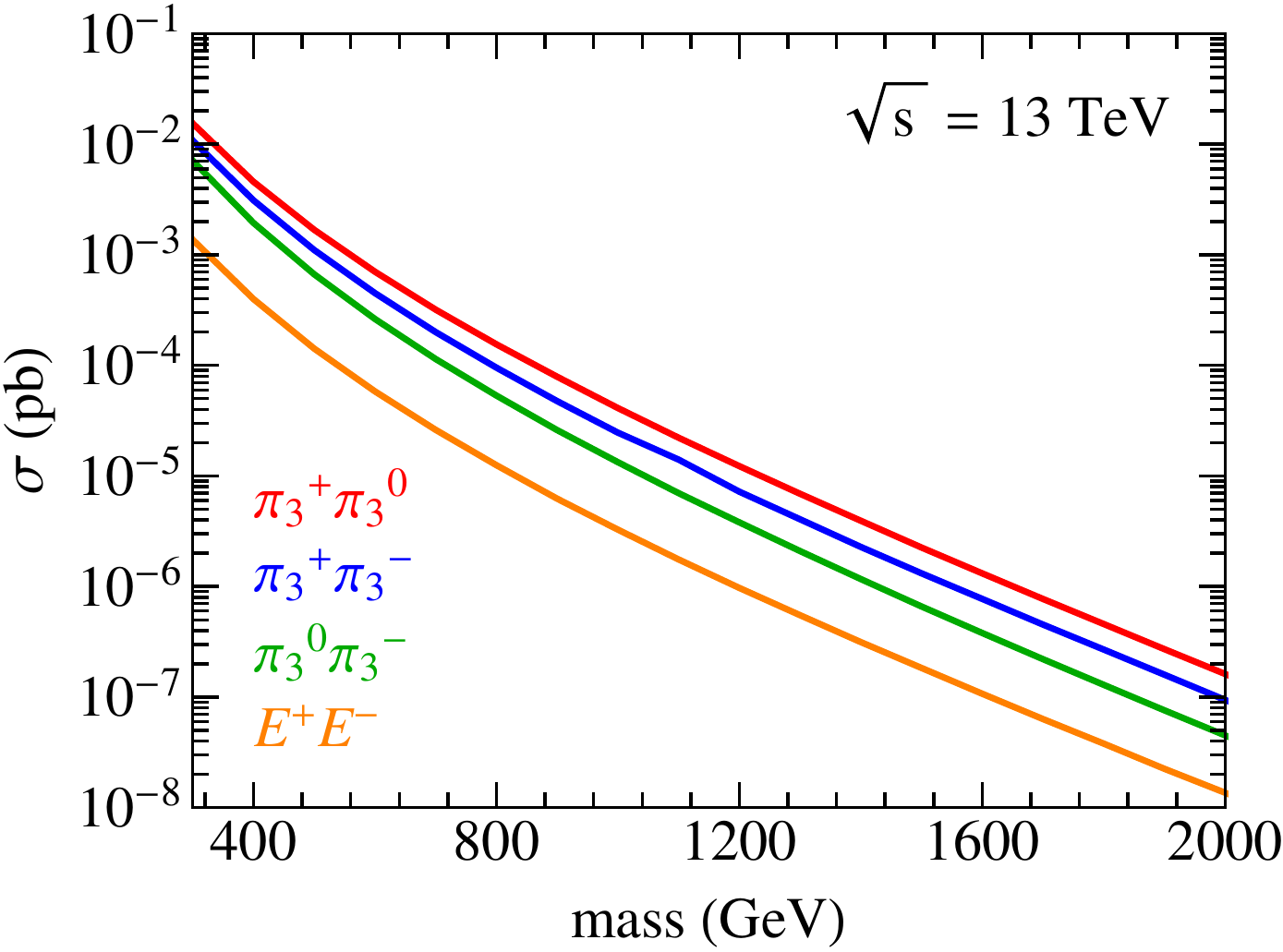}
\caption{The pair production cross section of colored pions (left) and of electroweakly charged pions (right) contained in the $\fourtyfour$ of Sp(10).}
\label{fig:pions24-xsec-pair}
\end{center}\end{figure}

In addition to these single production channels, both the real and complex pions can be pair produced at the LHC via their SM gauge interactions.  Clearly the colored pions, $\pi_8$, $Q_X$, $Q_Y$, and $Z_{\nicefrac{2}{3}}$, have the highest cross sections, shown in the left panel of Fig.~\ref{fig:pions24-xsec-pair}.  The smaller pair production cross sections for the electroweak states are shown in the right panel.  

Note that due to the QCD contribution to their masses, the colored pions are also heavier.  In addition to that the uncolored states can decay to final states with very low background, so it is not obvious \emph{a priori} what are the first signals that we should expect to observe. Before addressing this question in Sec.~\ref{sec:signals}, we move on to discussing the decays of the complex pions.

\subsubsection*{Complex Pion Decays for SU($N_F$) $\times$ SU($N_F$)} 

We first consider the SU(3)$_H$ and SU(4)$_H$ cases with $N_F = 5$, in which $Q_X$ is the only complex pion.  Given the interactions from the confining sector, $Q_X$ would be stable; however, it can decay through higher dimensional operators, just as charged pions in the SM decay through weak interactions.  Assuming baryon number $\bnum$ and lepton number $\lnum$ are good symmetries at the scale where the higher dimensional operators are generated, the decays depend on the $\bnum$ and $\lnum$ charges of the vector-like constituents of the pions.

With the ``usual'' assignment of $(\bnum,\lnum)=(\nicefrac{1}{3},0)$ for $D$ and $(\bnum,\lnum)=(0,1)$ for $L$ the leading decays come from
\begin{equation}\label{eq:lag-qx-usual}
\cL \supset c_1 \frac{f}{M_*^2} \partial_\mu Q_X \overline{\ell} \bar{\sigma}^\mu d^c
+ c_2 \frac{f}{M_*^2} \partial_\mu Q_X \overline{q} \bar{\sigma}^\mu e^c
\end{equation}
in which $M_*$ is the scale at which these operators are generated, and $c_i$ are numerical coefficients assumed to be $\cO(1)$.  Here the parametric dependence of the operators on $\Lambda$ and $f$ is estimated using NDA, in specific UV completions the operators coefficients can be further suppressed (for example by $(M_L+M_D)/\Lambda$, as we see in the following).\footnote{In order to keep the discussion as readable as possible, we only write down the leading operators.  When multiple operators related by equations of motion can be generated we typically write only one of them if including the others does not appreciably alter the branching ratios.  For instance, in Eq.~\eqref{eq:lag-qx-usual} the operators $Q_X d^c e^c H^c$, $Q_X u^c e^c H$, and $Q_X \overline{q} \overline{\ell} H$ are omitted because they lead to the same final state except in the case of subleading three body decays.}  As listed in Table~\ref{tab:decays-usual}, these interactions lead to the following decays of $Q_X = (X_{\nicefrac{-1}{3}}, X_{\nicefrac{-4}{3}})$:
\begin{subequations}\label{eq:qx-usual}\begin{align}
X_{\nicefrac{-1}{3}} & \to u \ell^-, d \nu, \\
X_{\nicefrac{-4}{3}} & \to d \ell^-.
\end{align}\end{subequations}
Here, and elsewhere, $u$ denotes any of the up-type quarks and $d$ denotes any of the down-type quarks.  Thus, although $Q_X$ decays as a leptoquark, squark searches can be sensitive to $X_{\nicefrac{-1}{3}}$ decays, due to the missing energy from the neutrino.  The decay length of $Q_X$ to fermions $a$ and $b$, with masses $m_a$ and $m_b$, is
\begin{equation} \label{eq:displaced}
\tau \simeq 0.1~{\rm mm}
\left(\frac{0.1}{c_i}\right)^2
\left(\frac{3~{\rm TeV}}{\Lambda}\right)^2
\left(\frac{M_*}{10~{\rm TeV}}\right)^4
\left(\frac{(1~{\rm GeV})^2}{m_a^2 + m_b^2}\right)
\left(\frac{1~{\rm TeV}}{M_\pi}\right).
\end{equation}
The helicity suppression leads to a preference for $Q_X$ to decay to third generation fermions.  

Since the operators in Eq.~\eqref{eq:lag-qx-usual} generate four-fermion interactions, they are constrained by flavor observables.  These limits can easily cause the $Q_X$ decays to become displaced. To see this we can consider a UV completion that generates the operators in Eq.~\eqref{eq:lag-qx-usual}. For example, consider adding a vector-like pair of chiral superfields, $\Phi$ and $\Phi^c$, which transform as a fundamental and antifundamental, respectively, of the confining group $G_H$, but as singlets of the flavor symmetry:
\begin{equation} \label{eq:superpotential1}
W \supset M_\Phi \Phi \Phi^c + \lambda_{L,i} \Phi^c L^c \ell_i + \lambda_{D,i} \Phi^c D d^c_i + M_L L L^c + M_D D D^c.
\end{equation}
Here $i$ is a SM flavor index and SU(5) enforces $\lambda_{D,i} = \lambda_{L,i}$ at the GUT scale.  These couplings are constrained by low-energy flavor observables, the most important of which are $K^0-\bar{K}^0$ oscillations and $\mu \rightarrow e \gamma$ decays.  The relevant effective operators are generated at one loop:
\begin{equation}\begin{aligned}
\mathcal{L} \supset \left(\frac{\lambda_{D,s} \lambda_{D,d}^*}{4 \pi M_\Phi}\right)^2 (\bar{d}^{c} \bar{\sigma}^{\mu} s^c)^2 
& + \frac{\lambda_{L,e} \lambda_{L,\mu}^* e}{16\pi^2} \frac{m_{\mu}}{M_\Phi^2} (\mu^c \sigma^{\mu \nu} e_L) F_{\mu \nu} + ... \\
\end{aligned}\end{equation}
dropping $\mathcal{O}(1)$ factors.  The MEG experiment places a bound on $\text{BR} (\mu \rightarrow e \gamma) < 5.7 \times 10^{-13}$~\cite{Adam:2013mnn}.  In terms of the singlet mass and couplings, this requires $M_\Phi^2 / (\lambda_{L,e} \lambda_{L,\mu}^*) \gtrsim (60\text{ TeV})^2$.  The absence of left-handed flavor violation in the quark sector weakens the limits from $K^0-\bar{K}^0$ oscillations~\cite{Isidori:2010kg}, for which $\Re{(M_{\Phi} / (\lambda_{D,s} \lambda_{D,d}^*))} \gtrsim 80 \text{ TeV}$.  The bound on CP violation, however, is somewhat more powerful, requiring $\Im{(M_{\Phi} / (\lambda_{D,s} \lambda_{D,d}^*))} \gtrsim 1.3\times 10^3 \text{ TeV}$.

With these restrictions, we can compute the minimum lifetime of $Q_X$.  The decay will proceed through the operators in Eq.~\eqref{eq:lag-qx-usual}, with $M_* = M_{\Phi}$ and $c_1 = \lambda_L^* \lambda_D (M_L + M_D) / \Lambda$.\footnote{Here we neglect the $(f / M_*^2) \partial_{\mu} Q_X \bar{q} \bar{\sigma}^{\mu} e^c$ operator as its coefficient is loop suppressed in this UV completion.} As long as the $\lambda$ couplings are sufficiently small, $\lambda \lesssim 0.05$, the dominant constraint comes from $\mu\to e \gamma$.  Assuming flavor anarchy in the $\Phi$ couplings, $Q_X$ decays must then be displaced.  For the parameters of Eq.~\eqref{eq:displaced}, we need $c_1 \lesssim 10^{-3}$ and thus $\tau \gtrsim 1$~m.  The next generation of lepton flavor violation experiments, such as MEG2~\cite{Baldini:2013ke} and Mu2e~\cite{Bartoszek:2014mya}, will improve limits on the mass scale of lepton flavor violation by almost an order of magnitude~\cite{deGouvea:2013zba}.  Thus, the observation of displaced $Q_X$ decays may predict measurable lepton flavor violation in the near future.

Other assignments of $\bnum$ and $\lnum$ are possible as well, for example the ``Higgs-like'' choice of $(\bnum,\lnum)=(0,0)$ for $L$, keeping the usual $(\bnum,\lnum)=(\nicefrac{1}{3},0)$ for $D$.  The leading operator in this case is
\begin{equation}\label{eq:lag-qx-higgslike}
\cL \supset 
c \frac{\Lambda f}{M_*^2} Q_X d^c \tilde{H}_u ,
\end{equation}
where $\tilde{H}_u$ is a higgsino, leading to squark-like decays of $Q_X$.  The decay is prompt, with
\begin{equation} \label{eq:prompt}
\tau \simeq 10^{-11}~{\rm m}
\left(\frac{0.1}{c}\right)^2
\left(\frac{3~{\rm TeV}}{\Lambda}\right)^4
\left(\frac{M_*}{10~{\rm TeV}}\right)^4
\left(\frac{1~{\rm TeV}}{M_\pi}\right).
\end{equation}
Alternatively, $(\bnum,\lnum)=(\nicefrac{-2}{3},0)$ for $D$ and $(\bnum,\lnum)=(0,0)$ for $L$ would allow the operator $\partial_\mu Q_X \bar{u^c} \bar{\sigma}^\mu q$.  For all other $\bnum$ and $\lnum$ choices $Q_X$ can decay only via operators of dimension higher than six, rendering it collider stable.  Moreover, only the first assignment, with a quark-like $D$ and a lepton-like $L$, has an SU(5)-invariant UV completion that contains no additional SM-charged matter and allows all of the pions (including the additional ones in the $\fourtyfour$ of Sp(10) discussed in the next section) to decay.

For $N_F > 5$, there are the additional complex pions $\phi_D$ and $\phi_L$.  Given the assignments $(\bnum,\lnum)=(\nicefrac{1}{3},0)$ for $D$, $(\bnum,\lnum)=(0,1)$ for $L$, and $(\bnum,\lnum)=(0,0)$ for $S$, the lowest dimensional operators are
\begin{equation}
\cL \supset c_L \frac{\Lambda f}{M_*^2} \phi_L e^c \tilde{H}_d + c_D \frac{\Lambda f}{M_*^2} \phi_D \bar{q} \bar{\tilde{H}}_d
\end{equation}
so that $\phi_D$ and $\phi_L$ decay like a squark and slepton, respectively.  However, with a ``Higgs-like'' assignment of $(\bnum,\lnum)=(\nicefrac{1}{3},0)$ for $D$ and $(\bnum,\lnum)=(0,0)$ for $L$, a new feature arises.  $\phi_L$ is now a scalar electroweak doublet without lepton number, and so in the presence of CP violation, the operator
\begin{equation}
\cL \supset c_{6H} \Lambda f \phi_L H,
\end{equation}
can be generated.  The $\phi_L$-Higgs mixing is interesting because it occurs even without electroweak symmetry breaking.  Of course, once electroweak symmetry is broken, the Higgs can mix with any of the singlets in the spectrum (provided that CP is broken) such as the $\pi_1$, $\pi'$, or $\pi_S$.  The mixing with these states, however, is induced by irrelevant operators such as
\begin{equation}\label{eq:higgs-mixing}
\cL \supset c \frac{\Lambda f}{M_*} \pi_S |H|^2\, .
\end{equation}

\begin{table}\begin{center}
\begin{tabular}{|cc|cc|cc|} \hline
\multicolumn{2}{|c|}{particle} & SU(5) $\times$ SU(5) & SU(10) & dominant decays & decay \\ \hline
$\pi_8$     & $(\mathbf{8},\mathbf{1})_0$                  & $\checkmark$ & $\checkmark$ & $gg, g\gamma, gZ$                       & prompt \\
$\pi_3^\pm$ & \multirow{2}{*}{$(\mathbf{1},\mathbf{3})_0$} & $\checkmark$ & $\checkmark$ & $W^\pm \gamma, W^\pm Z$                 & prompt \\
$\pi_3^0$   &                                              & $\checkmark$ & $\checkmark$ & $\gamma\gamma, Z\gamma, ZZ$             & prompt \\
$\pi_1$     & $(\mathbf{1},\mathbf{1})_0$                  & $\checkmark$ & $\checkmark$ & $gg, W^+W^-, ZZ, Z\gamma, \gamma\gamma$ & prompt \\
$\pi'$      & $(\mathbf{1},\mathbf{1})_0$                  & $\checkmark$ & $\checkmark$ & $gg, W^+W^-, ZZ, \gamma\gamma, Z\gamma$ & prompt \\ \hline
$X_{\nicefrac{-1}{3}}$ & \multirow{2}{*}{$(\mathbf{3},\mathbf{2})_{\nicefrac{-5}{6}}$} & $\checkmark$ & $-$ & $u\ell^-, d\nu$  & displaced, 3$^{\rm rd}$ gen \\
$X_{\nicefrac{-4}{3}}$ &                                                               & $\checkmark$ & $-$ & $d\ell^-$        & displaced, 3$^{\rm rd}$ gen \\
$X_{\nicefrac{-1}{3}}$ & \multirow{2}{*}{$(\mathbf{3},\mathbf{2})_{\nicefrac{-5}{6}}$} & $-$ & $\checkmark$ & $E^- d\ell^+$    & prompt \\
$X_{\nicefrac{-4}{3}}$ &                                                               & $-$ & $\checkmark$ & $E^- d\bar{\nu}$ & prompt \\
$Y_{\nicefrac{2}{3}}$  & \multirow{2}{*}{$(\mathbf{3},\mathbf{2})_{\nicefrac{1}{6}}$}  & $-$ & $\checkmark$ & $d\ell^+$        & prompt \\
$Y_{\nicefrac{-1}{3}}$ &                                                               & $-$ & $\checkmark$ & $d\bar{\nu}$     & prompt \\
$Z_{\nicefrac{2}{3}}$  & $(\mathbf{3},\mathbf{1})_{\nicefrac{2}{3}}$                   & $-$ & $\checkmark$ & $\bar{d_{[i}}\bar{d_{j]}}$ & prompt \\
$E^+$                  & $(\mathbf{1},\mathbf{1})_1$                                   & $-$ & $\checkmark$ & $\ell^+ \nu$     & prompt \\ \hline
\end{tabular}
\caption{Summary of pion decays with the usual $\bnum$ and $\lnum$ assignments ({\it i.e.}, $D$ quark-like and $L$ lepton-like). }
\label{tab:decays-usual}
\end{center}\end{table}

\subsubsection*{Complex Pion Decays for SU($2 N_F$)} 

We next consider the SU(2)$_H$ and Sp(4)$_H$ cases in which we have the full set of complex pions.  The additional states all decay through operators similar to that in Eq.~\eqref{eq:lag-qx-higgslike}, with $\bnum$ and $\lnum$ assignments determining the fermion flavor.  Their decays can therefore be prompt, with lifetimes given parametrically by Eq.~\eqref{eq:prompt}.  

For example, given the $\bnum$ and $\lnum$ assignment $(\nicefrac{1}{3},0)$ for $D$ and $(0,1)$ for $L$, the additional pions decay through
\begin{equation}\label{eq:44pionsUsualOps}
\cL \supset c_E \frac{\Lambda f}{M_*^2} E^+ \ell_{[i} \ell_{j]} + c_Y \frac{\Lambda f}{M_*^2} Q_Y d^c \ell + c_Z \frac{\Lambda f}{M_*^2} Z_{\nicefrac{2}{3}} d_{[i} d_{j]},
\end{equation}
in which the $i$ and $j$ indices indicate that the two fermions must have different flavor.  This leads to the final states of Table~\ref{tab:decays-usual}.  Note that as $E^+ \rightarrow \ell^+_{[i} \nu_{j]}$, the lepton flavor antisymmetry does not affect collider phenomenology.  For $Z_{\nicefrac{2}{3}}$, the flavor asymmetry is consequential: decays to $jj$ or to $bj$ are possible, but not $bb$.  Thus $Z_{\nicefrac{2}{3}}$ can appear as a stop with $R$-parity violation.  Cascades to other pions are possible, for example through $Z_{\nicefrac{2}{3}} \bar{d^c} \bar{\ell} Q_X$, $Z_{\nicefrac{2}{3}} q q E^-$, $Z_{\nicefrac{2}{3}} \bar{d^c} \bar{u^c} E^-$, or $Z_{\nicefrac{2}{3}} q \tilde{\lambda} Q_X$ (where $\tilde{\lambda}$ is a singlet fermion such as a bino or singlino), and generated at $\cO(1/M_*^2)$ by the strong dynamics.  However, due to the three-body phase space, these decays are subleading to $\bar{d}_{[i} \bar{d}_{j]}$.  Nevertheless, they could lead to more distinctive signals.

\begin{table}\begin{center}
\begin{tabular}{|cc|cc|cc|} \hline
\multicolumn{2}{|c|}{particle} & SU(5) $\times$ SU(5) & SU(10) & dominant decays & decay \\ \hline
$\pi_8$     & $(\mathbf{8},\mathbf{1})_0$                  & $\checkmark$ & $\checkmark$ & $gg, g\gamma, gZ$                       & prompt \\
$\pi_3^\pm$ & \multirow{2}{*}{$(\mathbf{1},\mathbf{3})_0$} & $\checkmark$ & $\checkmark$ & $W^\pm \gamma, W^\pm Z$                 & prompt \\
$\pi_3^0$   &                                              & $\checkmark$ & $\checkmark$ & $\gamma\gamma, Z\gamma, ZZ$             & prompt \\
$\pi_1$     & $(\mathbf{1},\mathbf{1})_0$                  & $\checkmark$ & $\checkmark$ & $gg, W^+W^-, ZZ, Z\gamma, \gamma\gamma$ & prompt \\
$\pi'$      & $(\mathbf{1},\mathbf{1})_0$                  & $\checkmark$ & $\checkmark$ & $gg, W^+W^-, ZZ, \gamma\gamma, Z\gamma$ & prompt \\\hline
$X_{\nicefrac{-1}{3}}$ & \multirow{2}{*}{$(\mathbf{3},\mathbf{2})_{\nicefrac{-5}{6}}$} & $\checkmark$ & $\checkmark$ & $d\tilde{\chi}^0$   & prompt \\
$X_{\nicefrac{-4}{3}}$ &                                                               & $\checkmark$ & $\checkmark$ & $d\tilde{\chi}^\pm$ & prompt \\
$Y_{\nicefrac{2}{3}}$  & \multirow{2}{*}{$(\mathbf{3},\mathbf{2})_{\nicefrac{1}{6}}$}  & $-$ & $\checkmark$ & $u\tilde{\chi}^0, d\tilde{\chi}^+$ & prompt \\
$Y_{\nicefrac{-1}{3}}$ &                                                               & $-$ & $\checkmark$ & $d\tilde{\chi}^0, u\tilde{\chi}^-$ & prompt \\
$Z_{\nicefrac{2}{3}}$  & $(\mathbf{3},\mathbf{1})_{\nicefrac{2}{3}}$                   & $-$ & $\checkmark$ & $\bar{d_{[i}}\bar{d_{j]}}$                   & prompt \\
$E^+$                  & $(\mathbf{1},\mathbf{1})_1$                                   & $-$ & $\checkmark$ & $u\bar{d}, \tilde{\chi}^+ \tilde{\chi}^0$ & prompt \\ \hline
\end{tabular}
\caption{Summary of pion decays with the Higgs-like $\bnum$ and $\lnum$ assignments ({\it i.e.}, $D$ quark-like and $L$ Higgs-like).}
\label{tab:decays-higgslike}
\end{center}\end{table}

Alternatively, Higgs-like $\bnum$ and $\lnum$ assignments give
\begin{equation}
\begin{aligned}
\cL \supset & \ c_{E1} \frac{f}{M_*^2} \partial_\mu E^+ \bar{d^c} \bar{\sigma}^\mu u^c
+ c_{E2} \frac{\Lambda f}{M_*^2} E^+ \tilde{H}_d \tilde{H}_d
+ c_{E3} \frac{\Lambda f}{M_*^2} E^+ \bar{\tilde{H}}_u \bar{\tilde{H}}_u \\
& +\ c_{Y1} \frac{\Lambda f}{M_*^2} Q_Y \bar{q} \bar{\tilde{\lambda}}
+ c_{Y2} \frac{\Lambda f}{M_*^2} Q_Y d^c \tilde{H}_d
+ c_{Y3} \frac{\Lambda f}{M_*^2} Q_Y u^c \tilde{H}_u
+ c_Z \frac{\Lambda f}{M_*^2} Z_{\nicefrac{2}{3}} d_{[i} d_{j]},
\end{aligned}
\end{equation}
with the leading decays given in Table~\ref{tab:decays-higgslike}.  When kinematically allowed, $E^\pm$ decays promptly to neutralinos and charginos; otherwise, the decay goes to $u\bar{d}$ and is displaced as a result of the chiral suppression.  The lifetime is then given by Eq.~\eqref{eq:displaced}.  $Q_Y$ decays are now squark-like, while $Z_{\nicefrac{2}{3}}$ decays through the same operator as before.

For $Q_X$, the operators in Eqs.~\eqref{eq:lag-qx-usual} and~\eqref{eq:lag-qx-higgslike} are still present for the appropriate $\bnum$ and $\lnum$ assignments.  Now, however, there are additional operators due to the presence of the other mesons.  For the standard $\bnum$ and $\lnum$ numbers, and given the four-fermion couplings that lead to Eq.~\eqref{eq:44pionsUsualOps}, the strong dynamics generates
\begin{equation}
\cL \supset c' \frac{\Lambda}{M_*^2} Q_X d^c \ell E^+.
\end{equation}
The $Q_X$ components then decay as
\begin{subequations}\begin{align}
X_{\nicefrac{-1}{3}} & \to d \ell^+ E^-    \to d \ell^+ (\ell^- \nu), \\
X_{\nicefrac{-4}{3}} & \to d \bar{\nu} E^- \to d \bar{\nu} (\ell^- \nu),
\end{align}\end{subequations}
with a lifetime
\begin{equation} \label{eq:cascade}
\tau \simeq
10^{-9}~{\rm m}
\left(\frac{0.1}{c'}\right)^2
\left(\frac{3~{\rm TeV}}{\Lambda}\right)^2
\left(\frac{M_*}{10~{\rm TeV}}\right)^4
\left(\frac{1~{\rm TeV}}{M_\pi}\right)^3.
\end{equation}
Similar cascades are not relevant for other pions that can decay directly to SM states without any chiral suppression.

In the simple UV completion of Eq.~\eqref{eq:superpotential1}, this cascade would not be generated, and the other complex pions in the $\fourtyfour$ would be stable due to unbroken, accidental symmetries.  In order to make these states decay, additional interactions are needed. Adding
\begin{equation} \label{eq:superpotential2}
W \supset \lambda_{L,i}^c \Phi L^c \ell_i + \lambda_{D,i}^c \Phi D d^c_i
\end{equation}
to the superpotential of Eq.~\eqref{eq:superpotential1} leads to the operators in Eq.~(\ref{eq:44pionsUsualOps}):
\begin{equation}
\mathcal{L} \supset \frac{\Lambda f b_{\Phi}^*}{M_{\Phi}^4} \left[\lambda_{L,i} \lambda_{L,j}^c E^+ \ell_{[i} \ell_{j]} + \lambda_{D,i} \lambda_{L,j}^c Q_Y d^c_i \ell_j + \lambda_{D,i} \lambda_{D,j}^c Z_{2/3}^* d^c_{[i} d^c_{j]}\right].
\end{equation}
Here $b_{\Phi}$ is the ($R$-symmetry breaking) soft mass for $\phi \phi^c$. To avoid flavor constraints, one only needs $c \sim \lambda^2 b_{\Phi}^* / M_{\Phi}^2 \lesssim 0.03$, so the decays of the complex pions can still be prompt.  This superpotential gives rise to the cascade decay for $Q_X$, with $c' \sim \lambda_D \lambda^c_L b_{\Phi}^* (M_L + M_D) / \Lambda M_{\Phi}^2 \lesssim 10^{-3}$.  Thus all of the pions in the {\bf 44} can have prompt decays, without tension with flavor observables. By contrast, if only the pions in the $\twentyfour$ are present, $Q_X$ decays are displaced (by at least 1 m).  There are other $\bnum$ and $\lnum$ choices that would allow all mesons to decay; however, each of these contains at least one meson that is collider stable.  We leave a detailed discussion of these more exotic assignments to future study.

\subsection{Vector Interactions}\label{sec:vectors}

Just as we did with the pions it is useful to sort the spin-1 mesons, the $\rho$'s, into two sets depending on their quantum numbers. We call the $\rho$'s that can mix with the SM gauge fields real vectors and the others complex vectors. Clearly the real vectors are much more relevant phenomenologically, since they can be singly produced at the LHC with a potentially large cross section.

The mixing between the real vectors and the SM gauge fields depends on the characteristic coupling strength of the strong sector $g_\rho$.  In the limit that $g_\rho$ is much larger than the SM gauge couplings $g_G$ the mixing is
\begin{equation}
\epsilon_G \simeq \frac{g_G}{g_\rho}.
\end{equation}
This induces interactions between the real vector $\rho_\mu$ and the SM current $J^\mu_G$ 
\begin{equation} \label{eq:vectorcurrent}
\cL \supset - \epsilon_G g_G \rho_\mu J^\mu_G.
\end{equation}
The effective coupling between real vectors and SM fermions is then $g_G^2/g_\rho$.  Thus, we can produce real vectors, including the electroweak triplet, from $q\bar{q}$ initial states.\footnote{The color octet vector can also couple to a pair of gluons through a dimension-6 operator.  If the coefficient is $\cO(1)$ the contribution to the production rate is comparable to that of $q\bar{q}$.  We neglect this contribution in the rest of the paper.}  Whether the interactions in Eq.~\eqref{eq:vectorcurrent} also give the main decay channels depends on the amount of chiral symmetry breaking.

If the pions are light, $2 m_\pi < m_\rho$, the vectors have an $\cO(1)$ width and decay dominantly to two pions.  Because of the large vector width this case results in an additional contribution to pion pair production with slightly different kinematics.  As the pions get heavier, $m_\rho/2 < m_\pi \lesssim m_\rho$, decays to two pions are kinematically forbidden and the dominant decays are either to a single pion and a SM vector through a dimension-5 operator or to a pair of SM fermions through the coupling in Eq.~\eqref{eq:vectorcurrent}.

The complex vectors, on the other hand, only couple to pairs of SM particles through higher dimensional operators analogously to the complex pions.  Since the scale at which these operators are generated is bounded by flavor constraints we expect pair production of complex vectors to dominate over single production.  Given the constraints on $\Lambda$ shown in Sec.~\ref{sec:signals}, it is unlikely that pair production of the complex $\rho$s will be relevant at the LHC.

\section{Signals} \label{sec:signals}

\subsection{Existing Direct Searches}

The models described so far have a variety of new particles that produce a wealth of new signals.  While the colored states are produced with larger cross sections, the new colorless states are lighter and lead to signatures with much lower backgrounds.  Therefore it is not immediately clear which signals we should expect to observe first at the LHC.  To develop some intuition, we compare the sensitivities of current analyses. 

\begin{figure}\begin{center}
\includegraphics[width=0.5\textwidth]{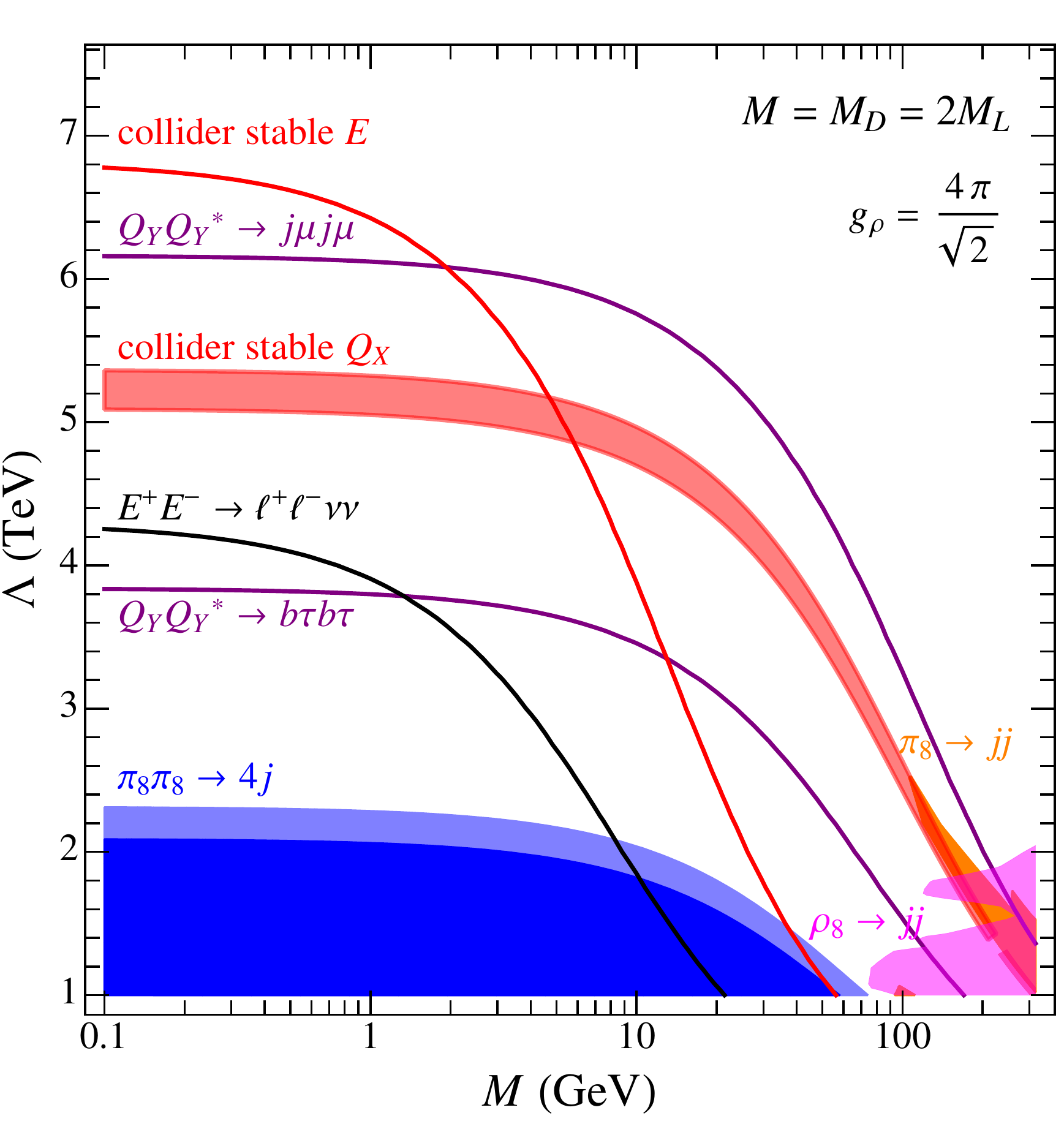}
\caption{Experimental constraints in the $\Lambda$ vs. $M$ plane, where $\Lambda$ is the confining scale and $M$ is the chiral symmetry breaking vector-like mass ($M = M_D = 2M_L$). We assume a single coupling for the strong sector, $g_\rho=4\pi/\sqrt{2}$.}
\label{fig:bounds}
\end{center}\end{figure}

In Fig.~\ref{fig:bounds} we show constraints from LHC searches.  If more than one search is sensitive to the same final state, we show and discuss only the strongest bound.  For each constraint, if branching ratios are not already fixed by anomalies, we assume a 100$\%$ branching ratio to the final state under consideration. We take $M_D=2 M_L$ from the assumption that the two parameters are equal at the GUT scale.  Furthermore, we assume that the strong sector is characterized by a single coupling which we fix to $g_\rho = 4\pi/\sqrt{2}$, as would be appropriate for $G_H =$ SU(2)$_H$.  With these two choices, only two parameters, $\Lambda$ and $M$ (where $M = M_D =2 M_L$), determine the decay and production rates of all the pions and vector mesons.  

The shaded regions in the figure correspond to constraints that cannot be evaded by judiciously choosing an assignment of baryon and lepton number for $D$ and $L$.  These regions pertain to the real vector mesons of the confining sector, as well as the real pions in the $\twentyfour$ of SU(5), which dominantly decay through the anomaly. 

Among the pions, the color octet gives the strongest of these constraints, both from its pair production and subsequent decays into jet pairs~\cite{Khachatryan:2014lpa} and from its single production~\cite{ATLAS:2015nsi, Aad:2014aqa, CMS:2015neg, Aaltonen:2008dn} (orange region in Fig.~\ref{fig:bounds}). In the case of pair production we show both the constraint from $\pi_8\to jj$ (darker blue) and the one from $\pi_8\to bj$ (lighter blue), all contained in~\cite{Khachatryan:2014lpa}.  Bounds from the other pions coupled to the anomalies are subdominant.  Therefore, when the complex pions decay into final states with low sensitivity (by an appropriate choice of the baryon and lepton numbers of the new quarks, the UV scale $M_*$, and the superpartner masses) $\pi_8$ is the first new particle expected to be observed at the LHC.

The complex pions become important, however, if $D$ has baryon number $\bnum = \nicefrac{1}{3}$, $L$ has lepton number $\lnum = 1$, and $M_*$ is low enough to allow prompt decays. In this case $Q_X$ and $Q_Y$ decay as leptoquarks and $E^\pm$ decays to a lepton and a neutrino. Depending on the value of the vector-like quark masses and the flavor structure of the model, we might either discover $E^\pm$ or one of the leptoquarks before $\pi_8$. This is shown in Fig.~\ref{fig:bounds} by means of two solid purple lines, below which we have an exclusion from current leptoquark searches, and a black solid line for $E^+\to \ell^+\nu$. The two choices of final state for the leptoquarks correspond to the most constraining ($\mu$+jet)~\cite{CMS:2016qhm} and least constraining ($\tau$+$b$)~\cite{Khachatryan:2014ura} possibilities.  The figure shows the exclusions for $Q_Y$, as these bounds apply to prompt decays, while in the simplest UV completions,  this assignment of baryon and lepton number causes $Q_X$ decays to be displaced.  The bounds on these displaced decays would be similar to the ones shown.\footnote{The sensitivity of current inclusive searches for displaced decays~\cite{CMS:2014hka, CMS:2014wda, Aad:2015rba, Aad:2015asa, Aad:2015uaa} depends on the lifetime of the particle. In the most favorable cases (cm $\lesssim \tau \lesssim$ m) it is comparable to that of the search for heavy stable charged particles shown in Fig.~\ref{fig:bounds}.}  The bound on $E^+\to \ell^+\nu$ omits different-flavor final states as well as those involving $\tau$s~\cite{Aad:2014vma}. LHC searches for decays to $\tau$s do not improve over the LEP bound in our case~\cite{Aad:2015eda, CMS:2016saj}.

If, however, $L$ does not carry lepton or baryon number, then $Q_X$ and $Q_Y$ decay as squarks (see Table~\ref{tab:decays-higgslike}), while $E^\pm$ decays to a chargino-neutralino pair.  These bounds can be made arbitrarily weak by an appropriate choice of gaugino or higgsino masses ({\it i.e.}, compressing the spectrum), and so are not shown.  The same logic applies to the additional states, $\phi_L$ and $\phi_D$, that are present in theories with $N_F > 5$.  It is worth mentioning, however, that if the lightest neutralino is massless, the bounds on the colored particles are comparable to the least sensitive one for leptoquarks ($\tau$+$b$ final states).  This gives roughly $m_{Q_{Y}} \gtrsim 600$~GeV, for decays to light quarks, and $m_{Q_{Y}} \gtrsim 800$~GeV, for decays to bottom or top quarks~\cite{CMS:2016wry, Khachatryan:2016xvy, CMS:2016nhb, ATLASsquarks, ATLASsbottoms}. The leptoquark exclusion, on the other hand, corresponds to $m_{Q_Y} \gtrsim 740$~GeV. 

The second case in which we would expect to see something before $\pi_8$ is if either $Q_X$, $Q_Y$ or $E^\pm$ are stable. The constraints from heavy stable charged particles searches~\cite{Chatrchyan:2013oca, CMS:2015kdx} are shown in the figure as solid red lines. The band associated to $Q_X$ depicts the uncertainty on the behavior of colored particles inside the LHC detectors: the top of the band corresponds to the CMS tracker + TOF analysis~\cite{CMS:2015kdx}, while the bottom corresponds to the CMS tracker-only analysis with the charge suppression model for the colored scalar propagation~\cite{CMS:2015kdx}.  At low $M$, these searches are more sensitive to $E^\pm$ than the other charged pions, even if $Q_X$ or $Q_Y$ are also stable. This is simply due to its smaller mass, $m_{E^\pm}/m_{Q} \sim g'/g_s + \cO(M/\Lambda)$. Inclusive searches for displaced decays~\cite{CMS:2014hka, CMS:2014wda, Aad:2015rba, Aad:2015asa, Aad:2015uaa} are at most as sensitive as the stable search for $Q_X$.

Finally, we briefly comment on the real vector mesons of the confining sector.  In most of the parameter space, they decay to a pair of pions and they are broad ($\Gamma/m=\mathcal{O}(1)$). In this case their main effect is to modify the constraints on the previously-discussed pions by contributing to their pair production cross section. We have not included this effect, which is typically negligible for the values of $\Lambda (\approx m_\rho)$ that we are considering.

On the other hand, when $m_\rho < 2 m_\pi$, the dominant decay of the real $\rho$s is to a pion and a SM gauge boson. In this case the $\rho$s are considerably narrower, and some of them have a significant branching ratio to pairs of SM fermions.  This leads, in the lower right corner of Fig.~\ref{fig:bounds}, to an approximately triangular excluded region whose vertices are $(M_1, \Lambda_1)\approx (60$~GeV, 1 TeV) and $(M_2, \Lambda_2)\approx (300$~GeV, 2 TeV). In this region, there is a bound from $\rho_8 \to jj$ that was computed assuming $\Lambda =m_\rho$. As with the other constraints, there are large uncertainties associated with the strong dynamics (particularly in this corner of parameter space where $M$ approaches $f$).

The pions that have not been mentioned explicitly, namely $\pi_1$, $\pi_3$, and $Z_{\nicefrac{2}{3}}$, give subdominant constraints (or equivalently are not candidates for early discovery), given current LHC searches. 

Before concluding, we note that when $N_F > 5$ and $L$ carries no lepton or baryon number, we can have mixing between the Higgs and $\phi_L$. Assuming that the mixing is small compared to the $\phi_L$ mass $m_{\phi_L}$, it scales as
\begin{equation}
\theta \sim \frac{c_{6H} \Lambda f}{m_{\phi_L}^2}\, ,
\end{equation}
in which $m_{\phi_L}^2=2(M_L+M_S) \Lambda +\mathcal{O}(\alpha_W \Lambda^2)$.  This is a CP violating effect, so there are constraints not only from Higgs coupling measurements and $\phi_L$ direct production, but also from electric dipole moment measurements.  Higgs coupling measurements alone, which require $\theta \lesssim 10\%$~\cite{Khachatryan:2014jba,Aad:2015gba}, rule out the entire $\Lambda$ vs. $M$ plane in Fig.~\ref{fig:bounds} if $M_S$ is comparable to $M_D$ and $M_L$ and $c_{6H}\sim \mathcal{O}(1)$.  In practice, this means that going beyond $N_F = 5$ requires either giving lepton number to $L$, imposing CP conservation in the UV, or adding large explicit chiral symmetry breaking. An alternate option is to take $c_{6H}$ small, which can be technically natural since it arises from the Yukawa coupling $L S H$ in the UV.

For $N_F\geq 7$ it is possible to have singlets that do not couple to the anomaly, but which mix with the Higgs after electroweak symmetry breaking. Their phenomenology is determined entirely from this mixing.

To summarize, if either $E^\pm$ or $Q_{X, Y}$ are stable, they give the first signal, with $E^\pm$ being the very first for $M\lesssim$~few GeV. Also, if $Q_{X, Y}$ decay as leptoquarks (either promptly or with some displacement) and $E^\pm$ decays to a first or second generation lepton plus a neutrino, we expect these three to be the first particles to be observed. In all other cases we will first see either $\pi_8$ or, for $M\gtrsim 60$~GeV and sufficiently small $\Lambda$, its vector partner $\rho_8$.  Dedicated searches for new signals that we describe in the next section can change this picture, making $\pi_3$ a candidate for early discovery. This, together with the bounds on $E^\pm$, demonstrates that we generically expect multiple states to be discovered almost concurrently, regardless of whether they are colored.

\subsection{Proposed Dedicated Searches}\label{sec:dsearches}

In addition to the signals described above, these models contain several new signals that the LHC collaborations are not currently looking for.  While inclusive searches have some sensitivity to these signals, dedicated searches would greatly improve the reach.

\paragraph{{\boldmath $Q_X$} cascade decays:} Because all of its decays to pairs of SM fermions are suppressed by Yukawas, it is quite plausible that 
\begin{equation}\begin{aligned}
X_{\nicefrac{-1}{3}} & \to d \ell^+ E^-    \to d \ell^+ (\ell^- \bar{\nu}), \\
X_{\nicefrac{-4}{3}} & \to d \bar{\nu} E^- \to d \bar{\nu} (\ell^- \bar{\nu})
\end{aligned}\end{equation}
have the largest branching ratio.

Furthermore, $Q_X$ is always pair produced, leading to final states with multiple leptons, jets, and missing energy.  These cascades can easily be prompt, as discussed in Sec.~\ref{sec:pheno}.  Distinguishing between these final states and the displaced decays of $Q_X$ can give information on the structure of the theory.  If the confining group is SU(2)$_H$ or Sp(4)$_H$, we expect cascade decays to dominate, while in the other cases (confining SU(3)$_H$ or SU(4)$_H$) $Q_X$ will appear as a long-lived leptoquark.  Of course, if $Q_X$ is collider stable, the two possibilities become degenerate.

\paragraph{{\boldmath $\pi_3\pi_3^*\to 4V$}:} The SU(2) triplet pion decays only to pairs of electroweak gauge bosons.  The neutral component decays to $\gamma\gamma$, $Z\gamma$, and $ZZ$, while the charged components decay to a $W$ along with a $\gamma$ or $Z$.  The branching ratios are shown in Fig.~\ref{fig:pions24-brs}.

Pair production of $\pi_3$ dominates over single production for masses accessible at the LHC, so we expect to observe final states with four gauge bosons.  A zero-background search in the 8 TeV dataset with a $10\%$ efficiency would be sensitive to $\pi_3$ masses up to $570$~GeV. However, this exclusion reach is reduced to $300$~GeV if removing backgrounds would require one of the $Z$ bosons to decay to leptons.  In the $\Lambda$ vs. $M$ plane of Fig.~\ref{fig:bounds} this corresponds to a sensitivity comparable to that of the search for a stable $Q_X$ or the $\pi_8$ paired dijet search, respectively. The 13 TeV dataset will become competitive once its integrated luminosity becomes comparable to that of the 8 TeV dataset. This shows once more that in these models both colored particles and states with only electroweak quantum numbers can be candidates for early discovery.

\paragraph{\boldmath $\pi_8\to gZ$:} The color octet pion can decay to a jet and a $Z$ boson, giving a nice resonant signal. As we discussed in the previous subsection, both single and pair production can be important, leading to a variety of final states: $(jj)(jZ)$, $(j\gamma)(jZ)$, $(jZ)(jZ)$, and $(jZ)$, in which the pairs in parentheses are resonant. While this is not a discovery channel, as the $gZ$ branching ratio is just a few permille, it would provide definitive evidence that the resonance in question is a $\pi_8$.  ATLAS and CMS are already looking for singly produced $j\gamma$ resonances~\cite{Aad:2015ywd, Aad:2013cva, Khachatryan:2014aka}, but all the other final states involving this decay are, as yet, unexplored.

\paragraph{Axion-like particles and other light bosons:} To conclude this section, we mention an interesting scenario that can arise naturally in our models. In the chiral limit, most of the phenomenology is unaltered; however, $\pi_1$ becomes almost massless, as remarked also in~\cite{Kilic:2009mi}. The dominant contribution to its mass then arises from QCD instanton effects:
\begin{equation}
m_{\pi_1} 
\approx \frac{N}{2\sqrt{15}}\frac{\Lambda_{\rm QCD}^2}{f} 
\approx 70~{\rm keV} \sqrt{\frac{N}{2}}  \left(\frac{5\;{\rm TeV}}{\Lambda}\right) ,
\quad\quad\quad 
g_\rho = 4\pi/\sqrt{N}.
\end{equation}
Its phenomenology is then equivalent to that of an axion with couplings to the gluon and electromagnetic field strengths of order
\begin{equation}
f_a^{G[\gamma]} \approx 1~{\rm TeV}\left[10^3~{\rm TeV}\right] \left(\frac{\Lambda}{5\; \rm TeV}\right) \sqrt{\frac{2}{N}}.
\end{equation}
While this possibility is already excluded by astrophysical observations and laboratory searches~\cite{Dobrich:2015jyk, Jaeckel:2015jla}, these bounds can be evaded for $m_{\pi_1} \gtrsim 400~{\rm MeV}$, which requires $M_D = 2M_L \gtrsim 10$~keV~$(5\;{\rm TeV}/\Lambda)$. 

Increasing the number of flavors increases the number of light particles. For example, theories with $N_F=6$ contain an extra singlet with couplings similar to $\pi_1$. Going beyond $N_F=6$ introduces an interesting feature: some of the new singlets might not couple to the anomaly, and their mixing with the Higgs in Eq.~(\ref{eq:higgs-mixing}) might be their only portal to the SM. This occurs if the mass matrix of the constituent fermions does not break all of the possible U(1)s.  Thus it is natural to expect axion-like particles, and possibly light Higgs-like particles, to appear in conjunction with the TeV scale physics discussed in the previous sections.

\subsection{Testing Unification}

If one were to observe the new pion states at the LHC, one of the first questions to ask would be whether the new matter is consistent with unification or not.  To this end, one can measure observables that are sensitive to the SU(5) symmetry.

SU(5) invariance (broken only by renormalization group running) predicts the ratio $M_D / M_L \approx 2$.  This prediction can be tested by measurements of the meson mass spectrum.  Assuming a common scale that cuts off the gauge contributions to the pion masses, one can write
\begin{equation}
\frac{M_D}{M_L} = \frac{6 g_3^2 \left(18 m_{Q_X}^2 - 9 m_{\pi_3}^2 - 8 m_{\pi_8}^2 \right) + \left(9 g_2^2 - 25 g'^2\right) m_{\pi_8}^2}{\left(6 g_3^2 - 25 g'^2\right) m_{\pi_3}^2 + 9 g_2^2 \left(8 m_{Q_X}^2 - 3 m_{\pi_3}^2 - 4 m_{\pi_8}^2\right)}.
\end{equation}
Here the gauge couplings should be evaluated at the cutoff scale, which is expected to be $\mathcal{O}(\Lambda)$.  In principle, this scale must itself be determined by further measurements of the strong sector; in practice, the scale dependence of the couplings is a small correction.

Furthermore, in the context of the explicit UV completion of Sec.~\ref{sec:PiPheno}, additional tests are possible.  For example, the displaced decays of $Q_X$ provide an opportunity to probe SU(5) invariance in the various branching fractions.  In appropriately constructed observables, such as
\begin{equation}
\frac{\text{BR}(Q_X \rightarrow \tau + j)}{\text{BR}(Q_X \rightarrow e + b) + \text{BR}(Q_X \rightarrow \mu + b)} \simeq \left(\frac{y_{\tau}}{y_b}\right)^2 \left(\frac{\lambda_{L,\tau}}{\lambda_{D,b}}\right)^2 \frac{\lambda_{D,d}^2 + \lambda_{D,s}^2}{\lambda_{L,e}^2 + \lambda_{L,\mu}^2},
\end{equation}
the effects of the strong coupling cancel, leaving only calculable SU(5) violation from the (weak) running of the couplings.

\section{Conclusions} \label{sec:conclusions}

In this work we consider extensions of minimal supersymmetry which preserve gauge coupling unification.  We have enumerated the list of viable theories with vector-like matter charged under a new gauge group.  In these theories, which happen to fall in the supersymmetric conformal window in the UV, the confinement scale is tied to the mass scale of the superpartners.  Minimal split supersymmetry places this confinement scale between $\sim 1 - 10^3$~TeV, so that the composite states of the new sector lie within reach of the LHC or, at worst, a 100 TeV collider.  

The low energy spectra of these theories include pion-like pseudoscalars and $\rho$-like vector mesons.  If explicit chiral symmetry breaking is small, we expect the pions to be the lightest new states. Their properties are summarized in Tables~\ref{tab:decays-usual} and~\ref{tab:decays-higgslike}.  While the precise phenomenology, including the pion lifetimes, depends on the details of the UV completion, limits on $\pi_8$ are unavoidable and force the confinement scale to be $\gtrsim$ 2 TeV.  If some of the colored or charged pions are collider stable, then the confinement scale is forced to be $\gtrsim$ 7 TeV, provided the pions are sufficiently light.

In addition, there are signals that would be interesting to explore further at the LHC: $Q_X$ cascade decays (involving final states with up to four leptons, two jets, and missing energy, with some of these objects being resonant), $jZ$ resonances (singly and pair produced and in association with $jj$ and $j\gamma$ resonances of the same mass), and final states with two pairs of resonant electroweak gauge bosons.  Furthermore, there is a natural limit of these theories in which TeV scale states are accompanied by ALPs and other light pseudoscalars whose properties can be predicted once enough is known about the spectrum of the heavier states.  If CP is violated, some of these states can mix with the SM Higgs and can be produced and decay in this way.

\section*{Acknowledgements}

The authors would like to thank Tom Banks, JiJi Fan, Philip Harris, Raman Sundrum, and Andrea Tesi for helpful discussions.  N. A.-H. is supported by the Department of Energy under grant number DE-FG02-91ER40654, R. T. D. is supported by a Marvin L. Goldberger Membership at the Institute for Advanced Study and DOE grant desc0009988, and M. L. is supported by a Peter Svennilson Membership at the Institute for Advanced Study.

\small
\bibliographystyle{utphys}
\bibliography{refs}

\providecommand{\href}[2]{#2}\begingroup\raggedright\begin{thebibliography}{10}

\bibitem{Dimopoulos:1981zb}
S.~Dimopoulos and H.~Georgi, ``{Softly Broken Supersymmetry and SU(5)},''
\href{http://dx.doi.org/10.1016/0550-3213(81)90522-8}{{\em Nucl. Phys.} {\bf
  B193} (1981)  150}.

\bibitem{Dienes:1996du}
K.~R. Dienes, ``{String theory and the path to unification: A Review of recent
  developments},'' \href{http://dx.doi.org/10.1016/S0370-1573(97)00009-4}{{\em
  Phys. Rept.} {\bf 287} (1997)  447--525},
\href{http://arxiv.org/abs/hep-th/9602045}{{\tt arXiv:hep-th/9602045
  [hep-th]}}.

\bibitem{Jungman:1995df}
G.~Jungman, M.~Kamionkowski, and K.~Griest, ``{Supersymmetric dark matter},''
  \href{http://dx.doi.org/10.1016/0370-1573(95)00058-5}{{\em Phys. Rept.} {\bf
  267} (1996)  195--373},
\href{http://arxiv.org/abs/hep-ph/9506380}{{\tt arXiv:hep-ph/9506380
  [hep-ph]}}.

\bibitem{Wells:2003tf}
J.~D. Wells, ``{Implications of supersymmetry breaking with a little hierarchy
  between gauginos and scalars},'' in {\em {11th International Conference on
  Supersymmetry and the Unification of Fundamental Interactions (SUSY 2003)
  Tucson, Arizona, June 5-10, 2003}}.
\newblock 2003.
\newblock
\href{http://arxiv.org/abs/hep-ph/0306127}{{\tt arXiv:hep-ph/0306127
  [hep-ph]}}.
\newblock

\bibitem{ArkaniHamed:2004fb}
N.~Arkani-Hamed and S.~Dimopoulos, ``{Supersymmetric unification without low
  energy supersymmetry and signatures for fine-tuning at the LHC},''
  \href{http://dx.doi.org/10.1088/1126-6708/2005/06/073}{{\em JHEP} {\bf 06}
  (2005)  073},
\href{http://arxiv.org/abs/hep-th/0405159}{{\tt arXiv:hep-th/0405159
  [hep-th]}}.

\bibitem{Giudice:2004tc}
G.~F. Giudice and A.~Romanino, ``{Split supersymmetry},''
  \href{http://dx.doi.org/10.1016/j.nuclphysb.2004.11.048,10.1016/j.nuclphysb.2004.08.001}{{\em
  Nucl. Phys.} {\bf B699} (2004)  65--89},
  \href{http://arxiv.org/abs/hep-ph/0406088}{{\tt arXiv:hep-ph/0406088
  [hep-ph]}}.
[Erratum: Nucl. Phys.B706,487(2005)].

\bibitem{ArkaniHamed:2006mb}
N.~Arkani-Hamed, A.~Delgado, and G.~F. Giudice, ``{The Well-tempered
  neutralino},'' \href{http://dx.doi.org/10.1016/j.nuclphysb.2006.02.010}{{\em
  Nucl. Phys.} {\bf B741} (2006)  108--130},
\href{http://arxiv.org/abs/hep-ph/0601041}{{\tt arXiv:hep-ph/0601041
  [hep-ph]}}.

\bibitem{Arvanitaki:2012ps}
A.~Arvanitaki, N.~Craig, S.~Dimopoulos, and G.~Villadoro, ``{Mini-Split},''
  \href{http://dx.doi.org/10.1007/JHEP02(2013)126}{{\em JHEP} {\bf 02} (2013)
  126},
\href{http://arxiv.org/abs/1210.0555}{{\tt arXiv:1210.0555 [hep-ph]}}.

\bibitem{ArkaniHamed:2012gw}
N.~Arkani-Hamed, A.~Gupta, D.~E. Kaplan, N.~Weiner, and T.~Zorawski, ``{Simply
  Unnatural Supersymmetry},''
\href{http://arxiv.org/abs/1212.6971}{{\tt arXiv:1212.6971 [hep-ph]}}.

\bibitem{ArkaniHamed:2004yi}
N.~Arkani-Hamed, S.~Dimopoulos, G.~F. Giudice, and A.~Romanino, ``{Aspects of
  split supersymmetry},''
  \href{http://dx.doi.org/10.1016/j.nuclphysb.2004.12.026}{{\em Nucl. Phys.}
  {\bf B709} (2005)  3--46},
\href{http://arxiv.org/abs/hep-ph/0409232}{{\tt arXiv:hep-ph/0409232
  [hep-ph]}}.

\bibitem{Banks:1993en}
T.~Banks, D.~B. Kaplan, and A.~E. Nelson, ``{Cosmological implications of
  dynamical supersymmetry breaking},''
  \href{http://dx.doi.org/10.1103/PhysRevD.49.779}{{\em Phys. Rev.} {\bf D49}
  (1994)  779--787},
\href{http://arxiv.org/abs/hep-ph/9308292}{{\tt arXiv:hep-ph/9308292
  [hep-ph]}}.

\bibitem{deCarlos:1993wie}
B.~de~Carlos, J.~A. Casas, F.~Quevedo, and E.~Roulet, ``{Model independent
  properties and cosmological implications of the dilaton and moduli sectors of
  4-d strings},'' \href{http://dx.doi.org/10.1016/0370-2693(93)91538-X}{{\em
  Phys. Lett.} {\bf B318} (1993)  447--456},
\href{http://arxiv.org/abs/hep-ph/9308325}{{\tt arXiv:hep-ph/9308325
  [hep-ph]}}.

\bibitem{Bagnaschi:2014rsa}
E.~Bagnaschi, G.~F. Giudice, P.~Slavich, and A.~Strumia, ``{Higgs Mass and
  Unnatural Supersymmetry},''
  \href{http://dx.doi.org/10.1007/JHEP09(2014)092}{{\em JHEP} {\bf 09} (2014)
  092},
\href{http://arxiv.org/abs/1407.4081}{{\tt arXiv:1407.4081 [hep-ph]}}.

\bibitem{Low:2014cba}
M.~Low and L.-T. Wang, ``{Neutralino dark matter at 14 TeV and 100 TeV},''
  \href{http://dx.doi.org/10.1007/JHEP08(2014)161}{{\em JHEP} {\bf 08} (2014)
  161},
\href{http://arxiv.org/abs/1404.0682}{{\tt arXiv:1404.0682 [hep-ph]}}.

\bibitem{Aguilar-Saavedra:2013qpa}
J.~A. Aguilar-Saavedra, R.~Benbrik, S.~Heinemeyer, and M.~Pérez-Victoria,
  ``{Handbook of vectorlike quarks: Mixing and single production},''
  \href{http://dx.doi.org/10.1103/PhysRevD.88.094010}{{\em Phys. Rev.} {\bf
  D88} (2013) no.~9, 094010},
\href{http://arxiv.org/abs/1306.0572}{{\tt arXiv:1306.0572 [hep-ph]}}.

\bibitem{Kilic:2009mi}
C.~Kilic, T.~Okui, and R.~Sundrum, ``{Vectorlike Confinement at the LHC},''
  \href{http://dx.doi.org/10.1007/JHEP02(2010)018}{{\em JHEP} {\bf 02} (2010)
  018},
\href{http://arxiv.org/abs/0906.0577}{{\tt arXiv:0906.0577 [hep-ph]}}.

\bibitem{Seiberg:1994pq}
N.~Seiberg, ``{Electric - magnetic duality in supersymmetric nonAbelian gauge
  theories},'' \href{http://dx.doi.org/10.1016/0550-3213(94)00023-8}{{\em Nucl.
  Phys.} {\bf B435} (1995)  129--146},
\href{http://arxiv.org/abs/hep-th/9411149}{{\tt arXiv:hep-th/9411149
  [hep-th]}}.

\bibitem{Intriligator:1995ne}
K.~A. Intriligator and P.~Pouliot, ``{Exact superpotentials, quantum vacua and
  duality in supersymmetric SP(N(c)) gauge theories},''
  \href{http://dx.doi.org/10.1016/0370-2693(95)00618-U}{{\em Phys. Lett.} {\bf
  B353} (1995)  471--476},
\href{http://arxiv.org/abs/hep-th/9505006}{{\tt arXiv:hep-th/9505006
  [hep-th]}}.

\bibitem{Masip:1998jc}
M.~Masip, R.~Munoz-Tapia, and A.~Pomarol, ``{Limits on the mass of the lightest
  Higgs in supersymmetric models},''
  \href{http://dx.doi.org/10.1103/PhysRevD.57.5340}{{\em Phys. Rev.} {\bf D57}
  (1998)  R5340},
\href{http://arxiv.org/abs/hep-ph/9801437}{{\tt arXiv:hep-ph/9801437
  [hep-ph]}}.

\bibitem{Jones:2008ib}
J.~L. Jones, ``{Gauge Coupling Unification in MSSM + 5 Flavors},''
  \href{http://dx.doi.org/10.1103/PhysRevD.79.075009}{{\em Phys. Rev.} {\bf
  D79} (2009)  075009},
\href{http://arxiv.org/abs/0812.2106}{{\tt arXiv:0812.2106 [hep-ph]}}.

\bibitem{Martin:2010kk}
S.~P. Martin, ``{Quirks in supersymmetry with gauge coupling unification},''
  \href{http://dx.doi.org/10.1103/PhysRevD.83.035019}{{\em Phys. Rev.} {\bf
  D83} (2011)  035019},
\href{http://arxiv.org/abs/1012.2072}{{\tt arXiv:1012.2072 [hep-ph]}}.

\bibitem{Karavirta:2011zg}
T.~Karavirta, J.~Rantaharju, K.~Rummukainen, and K.~Tuominen, ``{Determining
  the conformal window: SU(2) gauge theory with $N_f$ = 4, 6 and 10 fermion
  flavours},'' \href{http://dx.doi.org/10.1007/JHEP05(2012)003}{{\em JHEP} {\bf
  05} (2012)  003},
\href{http://arxiv.org/abs/1111.4104}{{\tt arXiv:1111.4104 [hep-lat]}}.

\bibitem{Fodor:2015baa}
Z.~Fodor, K.~Holland, J.~Kuti, S.~Mondal, D.~Nogradi, and C.~H. Wong, ``{The
  running coupling of 8 flavors and 3 colors},''
  \href{http://dx.doi.org/10.1007/JHEP06(2015)019}{{\em JHEP} {\bf 06} (2015)
  019},
\href{http://arxiv.org/abs/1503.01132}{{\tt arXiv:1503.01132 [hep-lat]}}.

\bibitem{Ryttov:2007sr}
T.~A. Ryttov and F.~Sannino, ``{Conformal Windows of SU(N) Gauge Theories,
  Higher Dimensional Representations and The Size of The Unparticle World},''
  \href{http://dx.doi.org/10.1103/PhysRevD.76.105004}{{\em Phys. Rev.} {\bf
  D76} (2007)  105004},
\href{http://arxiv.org/abs/0707.3166}{{\tt arXiv:0707.3166 [hep-th]}}.

\bibitem{Sannino:2009aw}
F.~Sannino, ``{Conformal Windows of SP(2N) and SO(N) Gauge Theories},''
  \href{http://dx.doi.org/10.1103/PhysRevD.79.096007}{{\em Phys. Rev.} {\bf
  D79} (2009)  096007},
\href{http://arxiv.org/abs/0902.3494}{{\tt arXiv:0902.3494 [hep-ph]}}.

\bibitem{Peskin:1980gc}
M.~E. Peskin, ``{The Alignment of the Vacuum in Theories of Technicolor},''
\href{http://dx.doi.org/10.1016/0550-3213(80)90051-6}{{\em Nucl. Phys.} {\bf
  B175} (1980)  197--233}.

\bibitem{Georgi:1992dw}
H.~Georgi, ``{Generalized dimensional analysis},''
  \href{http://dx.doi.org/10.1016/0370-2693(93)91728-6}{{\em Phys. Lett.} {\bf
  B298} (1993)  187--189},
\href{http://arxiv.org/abs/hep-ph/9207278}{{\tt arXiv:hep-ph/9207278
  [hep-ph]}}.

\bibitem{Adam:2013mnn}
{\bf MEG} Collaboration, J.~Adam {\em et al.}, ``{New constraint on the
  existence of the $\mu^+ \to e^+\gamma$ decay},''
  \href{http://dx.doi.org/10.1103/PhysRevLett.110.201801}{{\em Phys. Rev.
  Lett.} {\bf 110} (2013)  201801},
\href{http://arxiv.org/abs/1303.0754}{{\tt arXiv:1303.0754 [hep-ex]}}.

\bibitem{Isidori:2010kg}
G.~Isidori, Y.~Nir, and G.~Perez, ``{Flavor Physics Constraints for Physics
  Beyond the Standard Model},''
  \href{http://dx.doi.org/10.1146/annurev.nucl.012809.104534}{{\em Ann. Rev.
  Nucl. Part. Sci.} {\bf 60} (2010)  355},
\href{http://arxiv.org/abs/1002.0900}{{\tt arXiv:1002.0900 [hep-ph]}}.

\bibitem{Baldini:2013ke}
A.~M. Baldini {\em et al.}, ``{MEG Upgrade Proposal},''
\href{http://arxiv.org/abs/1301.7225}{{\tt arXiv:1301.7225 [physics.ins-det]}}.

\bibitem{Bartoszek:2014mya}
{\bf Mu2e} Collaboration, L.~Bartoszek {\em et al.}, ``{Mu2e Technical Design
  Report},''
\href{http://arxiv.org/abs/1501.05241}{{\tt arXiv:1501.05241
  [physics.ins-det]}}.

\bibitem{deGouvea:2013zba}
A.~de~Gouvea and P.~Vogel, ``{Lepton Flavor and Number Conservation, and
  Physics Beyond the Standard Model},''
  \href{http://dx.doi.org/10.1016/j.ppnp.2013.03.006}{{\em Prog. Part. Nucl.
  Phys.} {\bf 71} (2013)  75--92},
\href{http://arxiv.org/abs/1303.4097}{{\tt arXiv:1303.4097 [hep-ph]}}.

\bibitem{Khachatryan:2014lpa}
{\bf CMS} Collaboration, V.~Khachatryan {\em et al.}, ``{Search for
  pair-produced resonances decaying to jet pairs in proton–proton collisions
  at $\sqrt{s}$=8 TeV},''
  \href{http://dx.doi.org/10.1016/j.physletb.2015.04.045}{{\em Phys. Lett.}
  {\bf B747} (2015)  98--119},
\href{http://arxiv.org/abs/1412.7706}{{\tt arXiv:1412.7706 [hep-ex]}}.

\bibitem{ATLAS:2015nsi}
{\bf ATLAS} Collaboration, G.~Aad {\em et al.}, ``{Search for new phenomena in
  dijet mass and angular distributions from $pp$ collisions at $\sqrt{s}=$ 13
  TeV with the ATLAS detector},''
  \href{http://dx.doi.org/10.1016/j.physletb.2016.01.032}{{\em Phys. Lett.}
  {\bf B754} (2016)  302--322},
\href{http://arxiv.org/abs/1512.01530}{{\tt arXiv:1512.01530 [hep-ex]}}.

\bibitem{Aad:2014aqa}
{\bf ATLAS} Collaboration, G.~Aad {\em et al.}, ``{Search for new phenomena in
  the dijet mass distribution using $p-p$ collision data at $\sqrt{s}=8$ TeV
  with the ATLAS detector},''
  \href{http://dx.doi.org/10.1103/PhysRevD.91.052007}{{\em Phys. Rev.} {\bf
  D91} (2015) no.~5, 052007},
\href{http://arxiv.org/abs/1407.1376}{{\tt arXiv:1407.1376 [hep-ex]}}.

\bibitem{CMS:2015neg}
{\bf CMS} Collaboration, V.~Khachatryan {\em et al.},
``{Search for Resonances Decaying to Dijet Final States at $\sqrt{s} = 8$ TeV
  with Scouting Data},''.

\bibitem{Aaltonen:2008dn}
{\bf CDF} Collaboration, T.~Aaltonen {\em et al.}, ``{Search for new particles
  decaying into dijets in proton-antiproton collisions at s**(1/2) =
  1.96-TeV},'' \href{http://dx.doi.org/10.1103/PhysRevD.79.112002}{{\em Phys.
  Rev.} {\bf D79} (2009)  112002},
\href{http://arxiv.org/abs/0812.4036}{{\tt arXiv:0812.4036 [hep-ex]}}.

\bibitem{CMS:2016qhm}
{\bf CMS} Collaboration, V.~Khachatryan {\em et al.},
``{Search for pair-production of second-generation scalar leptoquarks in pp
  collisions at $\sqrt{s}=13~\mathrm{TeV}$ with the CMS detector},''.

\bibitem{Khachatryan:2014ura}
{\bf CMS} Collaboration, V.~Khachatryan {\em et al.}, ``{Search for pair
  production of third-generation scalar leptoquarks and top squarks in
  proton–proton collisions at $\sqrt{s}$=8 TeV},''
  \href{http://dx.doi.org/10.1016/j.physletb.2014.10.063}{{\em Phys. Lett.}
  {\bf B739} (2014)  229--249},
\href{http://arxiv.org/abs/1408.0806}{{\tt arXiv:1408.0806 [hep-ex]}}.

\bibitem{CMS:2014hka}
{\bf CMS} Collaboration, V.~Khachatryan {\em et al.}, ``{Search for long-lived
  particles that decay into final states containing two electrons or two muons
  in proton-proton collisions at $\sqrt{s} =$ 8 TeV},''
  \href{http://dx.doi.org/10.1103/PhysRevD.91.052012}{{\em Phys.Rev.} {\bf D91}
  (2015) no.~5, 052012},
\href{http://arxiv.org/abs/1411.6977}{{\tt arXiv:1411.6977 [hep-ex]}}.

\bibitem{CMS:2014wda}
{\bf CMS} Collaboration, V.~Khachatryan {\em et al.}, ``{Search for long-lived
  neutral particles decaying to quark-antiquark pairs in proton-proton
  collisions at $\sqrt{s} =$ 8 TeV},''
  \href{http://dx.doi.org/10.1103/PhysRevD.91.012007}{{\em Phys.Rev.} {\bf D91}
  (2015) no.~1, 012007},
\href{http://arxiv.org/abs/1411.6530}{{\tt arXiv:1411.6530 [hep-ex]}}.

\bibitem{Aad:2015rba}
{\bf ATLAS} Collaboration, G.~Aad {\em et al.}, ``{Search for massive,
  long-lived particles using multitrack displaced vertices or displaced lepton
  pairs in pp collisions at $\sqrt{s}$ = 8 TeV with the ATLAS detector},''
\href{http://arxiv.org/abs/1504.05162}{{\tt arXiv:1504.05162 [hep-ex]}}.

\bibitem{Aad:2015asa}
{\bf ATLAS} Collaboration, G.~Aad {\em et al.}, ``{Search for pair-produced
  long-lived neutral particles decaying in the ATLAS hadronic calorimeter in
  $pp$ collisions at $\sqrt{s}$ = 8 TeV},''
  \href{http://dx.doi.org/10.1016/j.physletb.2015.02.015}{{\em Phys.Lett.} {\bf
  B743} (2015)  15--34},
\href{http://arxiv.org/abs/1501.04020}{{\tt arXiv:1501.04020 [hep-ex]}}.

\bibitem{Aad:2015uaa}
{\bf ATLAS} Collaboration, G.~Aad {\em et al.}, ``{Search for long-lived,
  weakly interacting particles that decay to displaced hadronic jets in
  proton-proton collisions at $\sqrt{s}=8$ TeV with the ATLAS detector},''
\href{http://arxiv.org/abs/1504.03634}{{\tt arXiv:1504.03634 [hep-ex]}}.

\bibitem{Aad:2014vma}
{\bf ATLAS} Collaboration, G.~Aad {\em et al.}, ``{Search for direct production
  of charginos, neutralinos and sleptons in final states with two leptons and
  missing transverse momentum in $pp$ collisions at $\sqrt{s} =$ 8 TeV with the
  ATLAS detector},'' \href{http://dx.doi.org/10.1007/JHEP05(2014)071}{{\em
  JHEP} {\bf 05} (2014)  071},
\href{http://arxiv.org/abs/1403.5294}{{\tt arXiv:1403.5294 [hep-ex]}}.

\bibitem{Aad:2015eda}
{\bf ATLAS} Collaboration, G.~Aad {\em et al.}, ``{Search for the electroweak
  production of supersymmetric particles in $\sqrt{s}$=8 TeV $pp$ collisions
  with the ATLAS detector},''
  \href{http://dx.doi.org/10.1103/PhysRevD.93.052002}{{\em Phys. Rev.} {\bf
  D93} (2016) no.~5, 052002},
\href{http://arxiv.org/abs/1509.07152}{{\tt arXiv:1509.07152 [hep-ex]}}.

\bibitem{CMS:2016saj}
{\bf CMS} Collaboration, V.~Khachatryan {\em et al.},
``{Search for electroweak production of charginos in final states with two tau
  leptons in pp collisions at $\sqrt{s}=8~\mathrm{TeV}$},''.

\bibitem{CMS:2016wry}
{\bf CMS} Collaboration, V.~Khachatryan {\em et al.},
``{Search for direct production of bottom and light top squark pairs in
  proton-proton collisions at $\sqrt{s}=13~\mathrm{TeV}$},''.

\bibitem{Khachatryan:2016xvy}
{\bf CMS} Collaboration, V.~Khachatryan {\em et al.}, ``{Search for new physics
  with the MT2 variable in all-jets final states produced in pp collisions at
  sqrt(s) = 13 TeV},''
\href{http://arxiv.org/abs/1603.04053}{{\tt arXiv:1603.04053 [hep-ex]}}.

\bibitem{CMS:2016nhb}
{\bf CMS} Collaboration, V.~Khachatryan {\em et al.},
``{Search for direct production of top squark pairs decaying to all-hadronic
  final states in pp collisions at sqrt(s) = 13 TeV},''.

\bibitem{ATLASsquarks}
{\bf ATLAS} Collaboration, G.~Aad {\em et al.},
``{Search for squarks and gluinos in final states with jets and missing
  transverse momentum at $\sqrt{s}$ =13 TeV with the ATLAS detector},''.

\bibitem{ATLASsbottoms}
{\bf ATLAS} Collaboration, G.~Aad {\em et al.},
``{Search for Bottom Squark Pair Production with the ATLAS Detector in
  proton-proton Collisions at $\sqrt{s}=13$ TeV},''.

\bibitem{Chatrchyan:2013oca}
{\bf CMS} Collaboration, S.~Chatrchyan {\em et al.}, ``{Searches for long-lived
  charged particles in pp collisions at $\sqrt{s}$=7 and 8 TeV},''
  \href{http://dx.doi.org/10.1007/JHEP07(2013)122}{{\em JHEP} {\bf 1307} (2013)
   122},
\href{http://arxiv.org/abs/1305.0491}{{\tt arXiv:1305.0491 [hep-ex]}}.

\bibitem{CMS:2015kdx}
{\bf CMS} Collaboration, V.~Khachatryan {\em et al.},
``{Searchesrches for Long-lived Charged Particles in Proton-Proton Collisions
  at $\sqrt{s}=13$ TeV},''.

\bibitem{Khachatryan:2014jba}
{\bf CMS} Collaboration, V.~Khachatryan {\em et al.}, ``{Precise determination
  of the mass of the Higgs boson and tests of compatibility of its couplings
  with the standard model predictions using proton collisions at 7 and 8
  $\,\text {TeV}$},''
  \href{http://dx.doi.org/10.1140/epjc/s10052-015-3351-7}{{\em Eur. Phys. J.}
  {\bf C75} (2015) no.~5, 212},
\href{http://arxiv.org/abs/1412.8662}{{\tt arXiv:1412.8662 [hep-ex]}}.

\bibitem{Aad:2015gba}
{\bf ATLAS} Collaboration, G.~Aad {\em et al.}, ``{Measurements of the Higgs
  boson production and decay rates and coupling strengths using pp collision
  data at $\sqrt{s}=7$ and 8 TeV in the ATLAS experiment},''
  \href{http://dx.doi.org/10.1140/epjc/s10052-015-3769-y}{{\em Eur. Phys. J.}
  {\bf C76} (2016) no.~1, 6},
\href{http://arxiv.org/abs/1507.04548}{{\tt arXiv:1507.04548 [hep-ex]}}.

\bibitem{Aad:2015ywd}
{\bf ATLAS} Collaboration, G.~Aad {\em et al.}, ``{Search for new phenomena
  with photon+jet events in proton-proton collisions at $ \sqrt{s}=13 $ TeV
  with the ATLAS detector},''
  \href{http://dx.doi.org/10.1007/JHEP03(2016)041}{{\em JHEP} {\bf 03} (2016)
  041},
\href{http://arxiv.org/abs/1512.05910}{{\tt arXiv:1512.05910 [hep-ex]}}.

\bibitem{Aad:2013cva}
{\bf ATLAS} Collaboration, G.~Aad {\em et al.}, ``{Search for new phenomena in
  photon+jet events collected in proton--proton collisions at sqrt(s) = 8 TeV
  with the ATLAS detector},''
  \href{http://dx.doi.org/10.1016/j.physletb.2013.12.029}{{\em Phys. Lett.}
  {\bf B728} (2014)  562--578},
\href{http://arxiv.org/abs/1309.3230}{{\tt arXiv:1309.3230 [hep-ex]}}.

\bibitem{Khachatryan:2014aka}
{\bf CMS} Collaboration, V.~Khachatryan {\em et al.}, ``{Search for excited
  quarks in the $\gamma +$jet final state in proton–proton collisions at
  $\sqrt s=8$ TeV},''
  \href{http://dx.doi.org/10.1016/j.physletb.2014.09.048}{{\em Phys. Lett.}
  {\bf B738} (2014)  274--293},
\href{http://arxiv.org/abs/1406.5171}{{\tt arXiv:1406.5171 [hep-ex]}}.

\bibitem{Dobrich:2015jyk}
B.~Döbrich, J.~Jaeckel, F.~Kahlhoefer, A.~Ringwald, and K.~Schmidt-Hoberg,
  ``{ALPtraum: ALP production in proton beam dump experiments},''
  \href{http://dx.doi.org/10.1007/JHEP02(2016)018}{{\em JHEP} {\bf 02} (2016)
  018}, \href{http://arxiv.org/abs/1512.03069}{{\tt arXiv:1512.03069
  [hep-ph]}}.
[JHEP02,018(2016)].

\bibitem{Jaeckel:2015jla}
J.~Jaeckel and M.~Spannowsky, ``{Probing MeV to 90 GeV axion-like particles
  with LEP and LHC},''
  \href{http://dx.doi.org/10.1016/j.physletb.2015.12.037}{{\em Phys. Lett.}
  {\bf B753} (2016)  482--487},
\href{http://arxiv.org/abs/1509.00476}{{\tt arXiv:1509.00476 [hep-ph]}}.

\end{thebibliography}\endgroup
\end{document}